\documentclass[aps,pre,twocolumn,float]{revtex4-1}
\usepackage[normalem]{ulem}
\usepackage{amsmath}
\usepackage{amssymb}
\usepackage{color}
\usepackage{hyperref}
\usepackage{graphicx}

\DeclareMathAlphabet{\mathitbf}{OML}{cmm}{b}{it}

\renewcommand{\=}{\!=\!}
\newcommand{\ket}[1]{|#1\rangle}
\newcommand{\bra}[1]{\langle #1|}
\newcommand{\braket}[2]{\langle #1|#2\rangle}
\newcommand{\fv}{\mathitbf f}
\newcommand{\uv}{\mathitbf u}
\newcommand{\xv}{\mathitbf x}
\newcommand{\vv}{\mathitbf v}
\newcommand{\nv}{\mathitbf n}

\newcommand{\zerovector}{\mathBold 0}
\renewcommand{\th}{^{\mbox{\tiny th}}}

\newcommand{\mathBold}[1]{\mbox{\boldmath$#1$}}
\newcommand{\dbar}{{\,\mathchar'26\mkern-12mu d}}
\newcommand{\partialbar}{\partial\kern-0.5em\raise0.35ex\hbox{\footnotesize /}}

\newcommand{\sFrac}[2]{{\textstyle\frac{#1}{#2}}}

\definecolor{pink}{rgb}{1,1,0} 
\definecolor{red}{rgb}{1,0,0}
\definecolor{blue}{rgb}{0,0,1}
\definecolor{green}{rgb}{0,1,0}
\definecolor{yellow}{rgb}{1,1,0}
\definecolor{orange}{rgb}{1,0.5,0}
\definecolor{white}{rgb}{1,1,1}

\begin{document}

\title{Micromechanical theory of strain-stiffening of biopolymer networks} 

\author{Robbie Rens${}^{1}$, Carlos Villarroel${}^{2}$, Gustavo D\"uring${}^{2}$, and Edan Lerner${}^{1}$}
\affiliation{${}^1$Institute for Theoretical Physics, University of Amsterdam, Science Park 904, 1098 XH Amsterdam, The Netherlands \\ ${}^2$Instituto de F\'isica, Pontificia Universidad Cat\'olica de Chile, Casilla 306, Santiago, Chile}

\begin{abstract}
Filamentous bio-materials such as fibrin or collagen networks exhibit an enormous stiffening of their elastic moduli upon large deformations. This pronounced nonlinear behavior stems from a significant separation between the stiffnesses scales associated with bending vs.~stretching the material's constituent elements. Here we study a simple model of such materials --- floppy networks of hinged rigid bars embedded in an elastic matrix --- in which the effective ratio of bending to stretching stiffnesses vanishes identically. We introduce a theoretical framework and build upon it to construct a numerical method with which the model's micro- and macro-mechanics can be carefully studied. Our model, numerical method and theoretical framework allow us to robustly observe and fully understand the critical properties of the athermal strain-stiffening transition that underlies the nonlinear mechanical response of a broad class of biomaterials. 
\end{abstract}

\maketitle

\section{introduction}
 
Various types of biopolymers in biological systems form semi-flexible network structures. In mammals, examples are actin that spans parts of the intracellular cytoskeleton \cite{small1988actin,fletcher2010cell,gardel2004elastic}, collagen forms the extracellular matrix in which cells are embedded \cite{comper1996extracellular,hay1991cell}, and fibrin forms hemostatic clots \cite{weisel2004mechanical,weisel2007structure}.  Other, non-biological semi-flexible networks are also of interest, such as gels and network glasses \cite{bot1996large,bresser1986rigidity,thorpe2000self,thorpe2002generic}. Although various biopolymers can form networks radically different in their structure or microscopic interactions, their mechanical response shows a generic property \cite{storm2005nonlinear,C3SM50451D}. These materials are relatively soft at small deformations but stiff for larger deformations. This stiffening of the elastic moduli under external deformations, known as the strain-stiffening transition \cite{onck2005alternative,erk2010strain}, takes place abruptly at a finite strain, suggesting an underlying criticality.

At the origin of strain-stiffening transition is the existence of floppy modes in undercoordinated networks or \emph{frames} \cite{calladine1978buckminster}. The Maxwell criterion requires that for a frame of struts with freely hinged joints the average connectivity or \emph{coordination} $z$ must be larger than the critical value $z_c\=2\dbar$ in $\dbar$ spatial dimensions, in order to be mechanically rigid \cite{maxwell1864calculation}. In frames with connectivities lower than the critical connectivity, collective floppy modes emerge \cite{calladine1978buckminster,PhysRevLett.97.105501}, implying that these frames can be deformed while maintaining the invariance of all the struts' lengths. Floppy modes persist only for finite-amplitude deformations, above which the frame will eventually rigidify \cite{onck2005alternative,WeitzPRL2006,wyart2008elasticity}, at the strain at which no further deformation is possible without stretching or compressing the struts. 
 
Semi-flexible biopolymer networks are typically formed by cross-linking of fibers or fiber splitting, leading to a connectivity below the critical connectivity $z_c$. However, the joints of such networks are not freely hinged, indicating the presence of additional interactions. The latter give rise to macroscopic mechanical stability, as reflected by these systems' finite elastic moduli, despite the hypostaticity of the underlying network. The additional interactions that act as stabilizing fields could originate from pre-stress \cite{Gardel2006PNAS}, temperature fluctuations \cite{dennison2013fluctuation}, active stresses \cite{sheinman2012actively} or bending energy \cite{PhysRevB.47.703,rens2016nonlinear}. In this work we focus on the latter -- networks that are stabilized by bending energy, that arises due to the persistence of fibers. 

\begin{figure}[!ht]
\centering
\includegraphics[width = 0.5\textwidth]{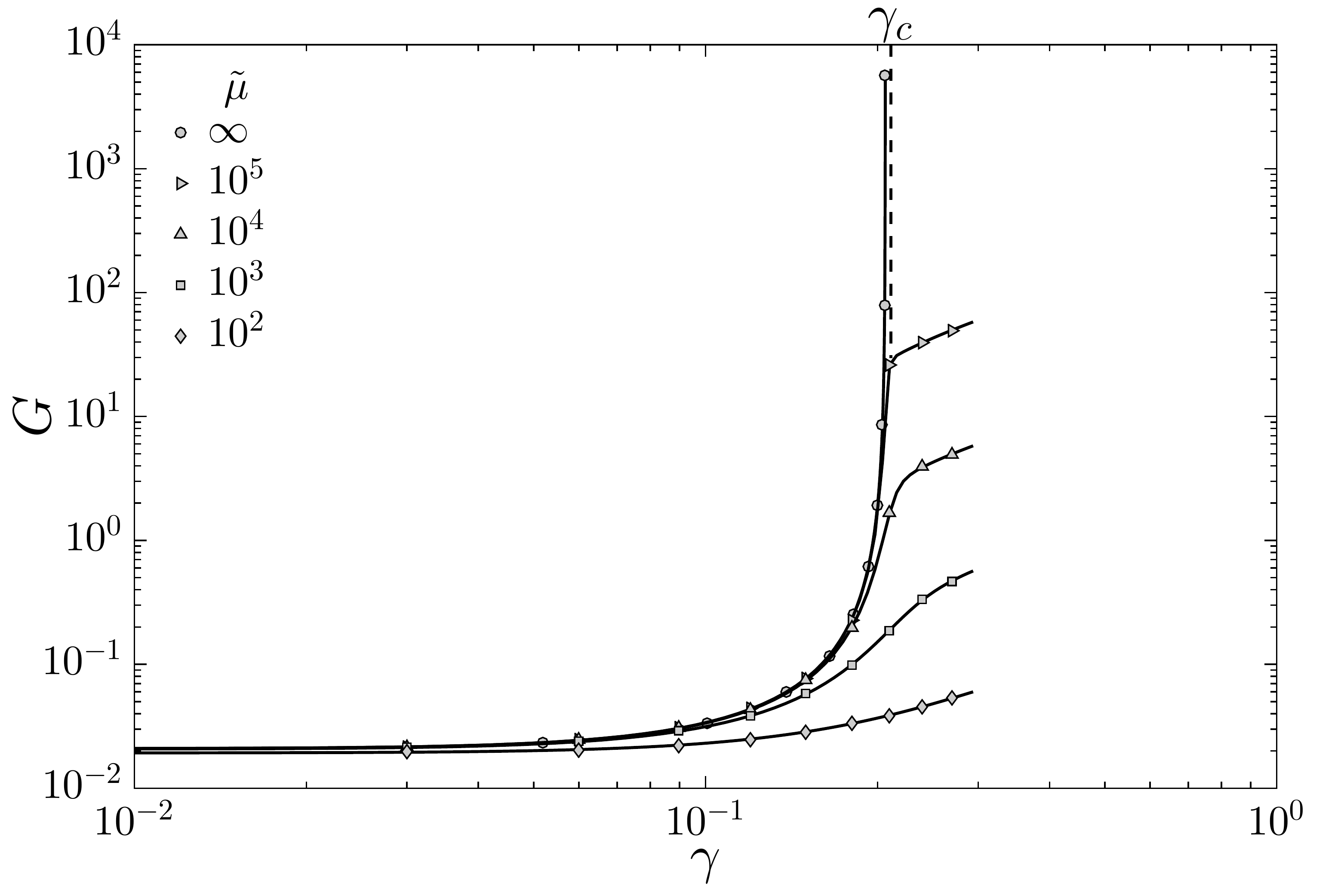}
\caption{\footnotesize The strain stiffening transition is conventionally studied by deforming simple model networks in which the characteristic energies of bending and stretching interactions are well-separated. In the figure we show the shear modulus $G$ of such a model (see text for details) as a function of shear strain, for different values of the ratio of stretching to bending stiffness $\tilde{\mu}$ (see Appendix~\ref{energeticmodel}). The curve with circle symbols corresponds to the shear modulus of the model introduced in this work, which corresponds to the limit $\tilde{\mu}\!\to\!\infty$. By studying this limit, our model allows us to reveal the critical properties of the strain stiffening transition that underlies the nonlinear mechanics of biopolymer networks.}
\label{model_motivation_fig}
\end{figure}

Biopolymer networks are often modelled in computational and theoretical studies by employing two characteristic energy scales, one  associated with the stretching of fibers and one associated with the stabilizing bending energy. When these two energy scales are well separated, a sharp stiffening transition upon imposing external deformation is observed \cite{broedersz2011criticality,sharma2016strain}, as demonstrated in Fig.~\ref{model_motivation_fig}. In \cite{sharma2016strain} it has been shown that the shear moduli of such networks follow a scaling form, suggesting a phase transition from a bending-dominated to a stretching-dominated regime. However, a thorough understanding of the micromechanics of this transition is still lacking.

In this work we introduce a theoretical framework that allows us to study in great detail the elastic properties of floppy frames of rigid struts stabilized by an embedding elastic energy. This model corresponds to an infinite scale separation between bending and stretching energies of conventional models of biopolymer networks, as demonstrated in Fig.~\ref{model_motivation_fig}. In turn, this allows us to cleanly reveal the critical behavior that underlies the strain-stiffening transition as seen in biomaterials \cite{motte2013strain}. Our framework allows us to derive an exact equation of motion that involves both the embedding elastic energy, and geometric information that characterizes the embedded floppy frames. We derive micromechanical expressions for elastic moduli, and perform a scaling analysis to determine the critical exponents that relate elastic moduli to deformation. We further resolve the mechanical dependence on our frames' connectivities, and identify a divergent micromechanical lengthscale that emerges upon approaching the strain stiffening transition. 

This work is organized as follows; in Sect.~\ref{theoretical_framework} we introduce our theoretical framework and use it to derive exact equations of motion and microscopic expressions of elastic moduli for externally-deformed frames of struts embedded in an elastic energy. In Sect.~\ref{numerics} we describe the numerical method that is designed based on our theoretical framework, and describe the protocols and numerical experiments we performed. Sect.~\ref{results} presents our theoretical analyses together with their validation by our numerical experiments. Concluding remarks are given in Sect.~\ref{conclusions}.


\section{Theoretical framework}
\label{theoretical_framework}

We consider disordered networks (frames) of rigid struts (edges) with mean connectivities $z$ smaller than the Maxwell threshold $z_c\=2\dbar$ ($\dbar$ denotes the dimension of space). The boundary conditions are assumed to be periodic in all dimensions, and Lees-Edwards boundary conditions \cite{allen1989computer} are employed for frames subjected to simple shear deformation. We consider the athermal limit, and neglect inertial effects, i.e.~the dynamics is overdamped. In addition to the geometric constraints imposed on the frame's $N$ nodes by its rigid edges, we embed our frame in an elastic medium by introducing a potential energy $U(\xv)$ that depends upon the $N\dbar$ coordinates $\xv$ of the frame's nodes. We do not specify at this point the properties of the potential energy function $U(\xv)$; these will however be discussed at a later stage in what follows. 

Our aim in this Section is three-fold; first, we derive the equations of motion that describe the microstructural evolution of our frames under externally-imposed, quasistatic deformations. We shall see that two sets of variables are key for resolving the microstructural evolution of our elastically-embedded frames: the deformation-induced nonaffine displacements of the frame's nodes, and the deformation-induced variation of tensile/contractile forces in the frame's edges. These variables are determined self-consistently in our framework by imposing two sets of contraints: the rigid-edge contraints that disallow variations in the distance between nodes connected by an edge, and the mechanical equilibrium contraints requiring that the forces exerted by the embedding potential energy must be balanced by the forces that arise in the rigid edges. We shall show that this requirement has important implications on the forces derived from the embedding potential energy. Finally, we derive expressions for elastic moduli at finite deformations, which are the main focus of our work. We note that the rate of imposed deformation is assumed to be very small, i.e.~timescales associated with the imposed deformation are assumed to be much slower than any other mechanical relaxation process in the system.

\subsection{Deformation-induced dynamics}
\label{node_dynamics}

Consider an elastically-embedded floppy frame under an external deformation; the latter is conventionally described in terms of an \emph{affine} transformation ${\cal H}(\gamma)$ parameterized by a single strain parameter $\gamma$; in the case of simple shear deformation in two-dimensions (2D) the affine transformation reads
\begin{equation}\label{foo43}
{\cal H}_{\mbox{\tiny simple shear}} = \left( \begin{array}{cc}1&\gamma\\0&1\end{array}\right)\,,
\end{equation}
whereas dilatational deformation results from imposing 
\begin{equation}\label{foo44}
{\cal H}_{\mbox{\tiny expansion}} = \left( \begin{array}{cc}1+\gamma&0\\0&1+\gamma\end{array}\right)\,.
\end{equation}
Deformation is imposed given ${\cal H}$ by applying the transformation $\xv\!\to\!{\cal H}(\gamma)\cdot\xv$ to the nodes' coordinates $\xv$.

In addition to the motion of the frame's nodes due to the imposed deformation as described by the affine transformations given above, the nodes must also perform additional, \emph{nonaffine displacements} $\delta \xv_k$, in order for the frame's perfectly-rigid edges to maintain their lengths. Here $\xv_k$ denotes the coordinates of the node with index~$k$. Under an infinitesimal strain $\delta\gamma$ the nonaffine displacements can be defined by the \emph{nonaffine velocities} (per unit strain) $\dot{\xv}_k$ as $\delta \xv_k\!=\!\dot{\xv}_k\delta \gamma$. The invariance of the length $r_{ij}\!\equiv\!|\xv_{ij}|$ of the edge connecting the $i\th$ and $j\th$ nodes under the imposed deformation can be expressed as
\begin{equation}\label{foo00}
\frac{dr_{ij}}{d\gamma} = \frac{\partialbar r_{ij}}{\partialbar\gamma} + \frac{\partial r_{ij}}{\partial\xv_\ell}\cdot\dot{\xv}_\ell = 0\,,
\end{equation}
where here and in what follows repeated indices are understood to be summed over, and we adopt a Lagrangian formulation in which the notation $\partialbar/\partialbar\gamma$ should be understood as the variation due to the imposed deformation, expressed in terms of the \emph{deformed} coordinates, see detailed discussion in Appendix~\ref{derivatives}; it reads
\begin{equation}
\frac{\partialbar}{\partialbar\gamma}  \equiv \sum_{i<j}\xv_{ij}\cdot\frac{d{\cal H}^T}{d \gamma}\cdot\frac{\partial}{\partial\xv_{ij}}\,\,.
\end{equation}

Since we consider undercoordinated frames with $z\!<\!z_c$, the set of geometric constraints embodied by Eq.~(\ref{foo00}) does not fully determine the nonaffine velocities $\dot{\xv}_k$, but rather merely constraints the space of possible nonaffine velocities. The full determination of $\dot{\xv}_k$ is possible by considering the consequences of \emph{mechanical equilibrium}, which we demand to hold under the imposed quasistatic deformation. Mechanical equilibrium can be expressed by considering the total net force $\fv_k$ on the $k\th$ node, which consists of two contributions: a contribution $-\partial U/\partial\xv_k$ from the embedding potential energy $U(\xv)$, and a contribution that arises from the the tensile or contractile forces $\tau_{ik}$ in the edges connected to the $k\th$ node. These two contributions must vanish in mechanical equilibrium, namely
\begin{equation}\label{foo01}
\fv_k = -\frac{\partial U}{\partial\xv_k} + \sum_{i(k)}\nv_{ik}\tau_{ik} = \zerovector\,,
\end{equation}
where the notation $i(k)$ is understood as the set of nodes $i$ that are connected to the $k\th$ node, $\nv_{ik}\!\equiv\!\xv_{ik}/r_{ik}$ is the unit vector that points from node $i$ to node $k$, and we chose the convention that $\tau_{ik}$ is positive for compressive forces. Eq.~(\ref{foo01}) is central to our theoretical framework; at every point along the deformation $\fv_k\=\zerovector$, which means that the frame's geometry, as encoded in the directors $\nv_{ik}$, and the tensile/contractile forces $\tau_{ik}$, will all evolve such that Eq.~(\ref{foo01}) is always satisfied. Notice that the edge forces $\tau_{ik}$ are still unspecified at this point, similarly to the nonaffine velocities $\dot{\xv}_k$, and will also be determined self-consistently.

Preservation of mechanical equilibrium under the imposed deformation means that the net forces $\fv_k$ on the nodes, which vanish identically as expressed by Eq.~(\ref{foo01}) above, remain unchanged, namely
\begin{eqnarray}
\frac{d\fv_k}{d\gamma} & = & \frac{\partialbar\fv_k}{\partialbar\gamma} + \frac{\partial\fv_k}{\partial\xv_\ell}\cdot\dot{\xv}_\ell +  \sum_{i(k)} \frac{\partial\fv_k}{\partial\tau_{ik}}\dot{\tau}_{ik} \nonumber \\
& = & -\frac{\partialbar}{\partialbar\gamma}\frac{\partial U}{\partial\xv_k} + \sum_{i(k)}\frac{\partialbar\nv_{ik}}{\partialbar\gamma}\tau_{ik} - \frac{\partial^2U}{\partial\xv_k\partial\xv_\ell}\cdot\dot{\xv}_\ell \nonumber \\
& & + \sum_{i(k)}\tau_{ik}\frac{\partial \nv_{ik}}{\partial\xv_\ell}\cdot\dot{\xv}_\ell+ \sum_{i(k)}\nv_{ik}\dot{\tau}_{ik}  = \zerovector\,. \label{foo02}
\end{eqnarray}
Eqs.~(\ref{foo00}) and (\ref{foo02}) form a closed linear system for the variables $\dot{\xv}_\ell$ and $\dot{\tau}_{ik}\!\equiv\! d\tau_{ik}/d\gamma$, that can be written in matrix, bra-ket form as 
\begin{equation}\label{foo03}
\left( \begin{array}{cc}{\cal A}(\tau)&-{\cal S}^T\\-{\cal S}&0\end{array}\right)
\left( \begin{array}{c}\ket{\dot{\xv}}\\\ket{\dot{\tau}}\end{array}\right) = 
\left( \begin{array}{c}\ket{\partialbar_\gamma\fv(\tau)} \\ \ket{\partialbar_\gamma r}\end{array}\right)\,,
\end{equation}
where we have defined
\begin{eqnarray}
{\cal A}_{k\ell}(\tau) & \equiv & \frac{\partial \fv_k}{\partial\xv_\ell} =  \frac{\partial^2U}{\partial\xv_k\partial\xv_\ell} - \sum_{i(k)}\tau_{ik}\frac{\partial \nv_{ik}}{\partial\xv_\ell}\,, \label{foo13} \\
{\cal S}_{ij,k} & \equiv & \frac{\partial r_{ij}}{\partial \xv_k}\,, \label{es_definition} \\ 
\partialbar_\gamma \fv_k(\tau) & \equiv & \frac{\partialbar \fv_k}{\partialbar\gamma} = \sum_{i(k)}\frac{\partial\nv_{ik}}{\partialbar\gamma}\tau_{ik}-\frac{\partialbar}{\partialbar\gamma}\frac{\partial U}{\partial\xv_k} \,, \label{foo20} \\
\partialbar_\gamma r_{ij} & \equiv & \frac{\partialbar r_{ij}}{\partialbar\gamma}\,. \label{foo14}
\end{eqnarray} 

Eq.~(\ref{foo03}) uniquely determines the dynamics of the system under any imposed deformation, parameterized here by the strain parameter $\gamma$. The notations ${\cal A}(\tau)$ and $\partialbar_\gamma \fv_k(\tau)$ are meant to emphasize that these objects depend on the \emph{set} of $Nz/2$ edge forces $\tau_{ik}$, the latter are discussed in length in Subsection~\ref{edge_forces} below. Notice that $\ket{\partialbar_\gamma\fv}$ is understood to represent a node-wise vector with $N\!\times\!\dbar$ components, whereas $\ket{\partialbar_\gamma r}$ is understood to represent an edge-wise vector with $Nz/2$ components. Importantly, we will assume that undeformed frames (i.e.~prior to any applied deformation) are unstressed, i.e.~their edges carry no initial tensile or compressive forces, meaning that $\tau_{ij}|_{\gamma\=0}\=0$ for all edges $ij$, and that $\partial U/\partial\xv_k |_{\gamma\=0}\=\zerovector$. Consequently, ${\cal A}_{k\ell}$ reduces to the Hessian matrix of the elastic energy $\frac{\partial^2U}{\partial\xv_k\partial\xv_\ell}$ in undeformed frames, as understood from Eq.~(\ref{foo13}). 

We further highlight that the operator ${\cal S}$ is known as the \emph{equilibrium matrix} \cite{calladine1978buckminster}; it holds geometric information of the frame's rigid edges, and plays an important role in determining the mechanics and rheology of floppy systems close to the jamming point \cite{asm_pnas,during2013phonon, asm_strain_stiffening, ASM_2016}, as well as the elasticity of random networks of Hookean springs \cite{breakdown}, and the physics of topological metamaterials \cite{Kane2013}. In our rigid-edge floppy frames, ${\cal S}$ has a nonzero kernel, i.e.~there exist nontrivial displacement fields on the frame's nodes, that --- to linear order in the displacements magnitude --- preserve the rigid-edge constraints. Such displacements that neither stretch nor compress the edges are conventionally termed \emph{floppy modes}, and will be further discussed below. 

We note that the existence of a unique solution to Eq.~(\ref{foo03}) depends on the number of interactions that the embedding potential energy $U$ is comprised of, that are not redundant with respect to the constraints embodied by the rigid-edge network (e.g., an interaction between a pair of nodes that are already connected by a rigid edge is redundant). The system will possess finite elastic moduli only if the number of non-redundant interactions is larger or equal to $N(z_c\!-\! z)$; in what follows we consider embedding potential energies that comprise of many more non-redundant interactions compared to this bound, which guarantees the existence of a unique solution to Eq.~(\ref{foo03}).

\subsection{Edge tensile/compressive forces}
\label{edge_forces}

The deformation-induced variations $d\tau_{ij}/d\gamma$ are determined by Eq.~(\ref{foo03}) (denoted there by $\ket{\dot{\tau}}$). The edge forces $\ket{\tau}$ at strains $\gamma\!>\!0$ are therefore given by the integrals $\tau_{ij}\!=\!\int_0^\gamma (d\tau_{ij}/d\gamma')d\gamma'$, with the initial conditions $\tau_{ij}(\gamma\!=\!0)\=0$ for pairs of nodes $i,j$ connected by a rigid edge.  Equivalently, an expression for the edge forces can be obtained by considering the bra-ket form of Eq.~(\ref{foo01}), namely
\begin{equation}\label{foo04}
\ket{\fv} = {\cal S}^T\ket{\tau} - \ket{\partial_\xv U} = \zerovector\,,
\end{equation}
where $\ket{\partial_\xv U} \!\equiv\! \ket{\partial U/\partial\xv}$.

Eq.~(\ref{foo04}) highlights an important property of the potential-derived forces $\ket{\partial_\xv U}$; this can be seen by considering any floppy mode $\ket{\uv}$, i.e.~any displacement field on the nodes on the frame that, to linear order in the field's magnitude, does not violate the frame's rigid-edge constraints. Floppy modes $\ket{\uv}$ form the kernel of the operator ${\cal S}$ (defined in Eq.~(\ref{es_definition})), i.e.~they satisfy the relation ${\cal S}\ket{\uv}\!=\!0$. Contracting $\bra{\uv}$ with Eq.~(\ref{foo04}), we obtain
\begin{equation}\label{foo06}
\bra{\uv}{\cal S}^T\ket{\tau} - \braket{\uv}{\partial_\xv U} = - \braket{\uv}{\partial_\xv U} = 0\,.
\end{equation}
We therefore conclude that the dynamics drives the system in a very particular way: the requirement that the potential derived forces $\ket{\partial_\xv U}$ must be exactly balanced by the edge forces $\ket{\tau}$ at any point along the deformation implies that the frame's nodes \emph{self-organize} under the imposed deformation such that $\ket{\partial_\xv U}$ acquires no projection onto the floppy modes of the rigid-edge frame.

This property of the potential-derived forces $\ket{\partial_\xv U}$ as expressed by Eq.~(\ref{foo06}) is discussed further in Appendix~\ref{projection_operator_appendix}; it allows us to write an explicit expression for the edge forces $\ket{\tau}$, that will be key for our scaling analysis later on. This is done by operating on Eq.~(\ref{foo04}) with ${\cal S}$ and rearranging it in favor of the edge forces $\ket{\tau}$ as
\begin{equation}\label{foo08}
\ket{\tau} = ({\cal S}{\cal S}^T)^{-1}{\cal S}\ket{\partial_\xv U}\,.
\end{equation}
The zero projection of $\ket{\partial_\xv U}$ on the kernel of ${\cal S}$, as expressed by Eq.~(\ref{foo06}), guarentees that Eq.~(\ref{foo08}) for the edge forces is a solution to the mechanical equilibrium equation (\ref{foo04}).

\subsection{Explicit equations of motion}

The availability of an explicit expression for the edge forces as given by Eq.~(\ref{foo08}) allows us to express the equations of motion for the nodes and the edge forces as \emph{explicit functions of a given state}; some algebraic manipulations of Eqs.~(\ref{foo03}) and the adoption of bra-ket notation for the sake of clarity, yields
\begin{eqnarray}
\ket{\dot{\tau}} & = & -\big({\cal S}{\cal A}^{-1}{\cal S}^T\big)^{-1}\big( \ket{\partialbar_\gamma r} + {\cal S}{\cal A}^{-1}\ket{\partialbar_\gamma\fv} \big)\,, \label{foo15} \\
\ket{\dot{\xv}} & = & {\cal A}^{-1}\ket{\partialbar_\gamma \fv} \nonumber\\
&& \!\!\!-{\cal A}^{-1}{\cal S}^T\!\big({\cal S}{\cal A}^{-1}{\cal S}^T\big)^{-1}\!\big( \ket{\partialbar_\gamma r}\!+\!{\cal S}{\cal A}^{-1}\ket{\partialbar_\gamma\fv} \big) \,. \label{foo16}
\end{eqnarray}
As discussed above and seen in Eqs.~(\ref{foo13})-(\ref{foo14}), here and in what follows ${\cal A}$ and $\ket{\partialbar_\gamma \fv}$ are understood to depend on the set of $Nz/2$ edge forces $\ket{\tau}$, the latter are explicitly given by Eq.~(\ref{foo08}). If ${\cal A}$ possesses zero modes, ${\cal A}^{-1}$ in the above relations should be understood as representing the pseudo inverse, as implied by Eq.~(\ref{foo03}). Eqs.~(\ref{foo15}) and (\ref{foo16}) will be employed in our scaling analysis in what follows.

\subsection{Elastic moduli}
\label{elastic_moduli}

The deformation dynamics of our elastically-embedded frames is fully described by Eqs.~(\ref{foo15}) and (\ref{foo16}). We now turn to deriving microscopic expressions for elastic moduli. Recall that we consider the athermal limit, then elastic moduli $E$ are defined as 
\begin{equation}\label{foo12}
E = \frac{1}{V}\frac{d^2U}{d\gamma^2}\,,
\end{equation}
with $V\!=\! L^\dbar$ denoting the system's volume. Full derivatives $d/d\gamma$ are understood as taken under two sets of constraints: the rigid-edge constraints that disallow the frame's edges to change their length (as expressed by Eq.~(\ref{foo00})), and the mechanical equilibrium constraints that imply that edge tensile/compressive forces are always balanced by the potential-derived forces (as expressed by Eqs.~(\ref{foo02}) and (\ref{foo04})). Specifically, full derivatives read
\begin{equation}
\frac{d}{d\gamma} = \frac{\partialbar}{\partialbar\gamma} + \dot{\xv}_k\cdot \frac{\partial}{\partial\xv_k} + \sum_{\mbox{\tiny edges }ij}\dot{\tau}_{ij} \frac{\partial}{\partial \tau_{ij}}\,,
\end{equation}
as employed, e.g., in Eq.~(\ref{foo02}). In what follows we will denote by $G$ the shear modulus, obtained when $\gamma$ represents simple shear strain, and by $K$ the bulk modulus, obtained when $\gamma$ represents compressive/dilatational strain.

We start our derivation of microscopic expressions for elastic moduli of elastically-embedded floppy frames with 
\begin{equation}\label{foo09}
dU/d\gamma = \partialbar U/\partialbar\gamma + \braket{\partial_{\xv} U}{\dot{\xv}}\,.
\end{equation}
In generic athermal disordered solids the potential-derived forces $\ket{\partial_{\xv} U}$ vanish by virtue of mechanical equilibrium, leading to the vanishing of the second term on the right-hand side (RHS) of Eq.~(\ref{foo09}), see e.g.~\cite{lutsko}. Importantly, in our framework we also assume mechanical equilibrium; however, as discussed in length above, it emerges due to the \emph{balance} between the edge forces $\ket{\tau}$ and the potential-derived forces $\ket{\partial_{\xv} U}$ (see Eq.~(\ref{foo04})), which are each generally non-zero. Consequently, the second term on the RHS of Eq.~(\ref{foo09}) does not vanish, nor does its derivative with respect to strain, and therefore the second full derivative of the potential energy with respect to strain reads

\begin{eqnarray}\label{foo09B}
d^2U/d\gamma^2 &=& \partialbar^2U/\partialbar\gamma^2+\braket{\partial_\xv\partialbar_\gamma U}{\dot{\xv}}+\braket{\partialbar_{\gamma}\partial_\xv U}{\dot{\xv}}\nonumber\\
&&+\  \bra{\dot{\xv}}{\cal M}\ket{\dot{\xv}}+\braket{\partial_\xv U}{\ddot{\xv}}\,,
\end{eqnarray}
where we utilized the conventional notation for the Hessian matrix of the potential energy ${\cal M}\!\equiv\!\frac{\partial^2U}{\partial\xv\partial\xv}$, and see discussion about the non-commuting variations $\partial_\xv\partialbar_\gamma\!\ne\!\partialbar_\gamma\partial_\xv$ in Appendix~\ref{derivatives}). Importantly, we note that the RHS of the above equation depends on the still-undetermined variables $\ket{\ddot{\xv}}\!\equiv\! \ket{d\dot{\xv}/d\gamma}$.  

Instead of deriving an equation for $\ket{\ddot{\xv}}$ (which is possible but tedious), we show next that elastic moduli can be expressed solely in terms of previously-determined quantities. First, the force balance equation (\ref{foo04}) can be used to write the second term of the RHS of Eq.~(\ref{foo09}) as
\begin{equation}\label{foo10}
\braket{\partial_\xv U}{\dot{\xv}} = \bra{\tau}{\cal S}\ket{\dot{\xv}} = -\braket{\tau}{\partialbar_\gamma r}\,,
\end{equation}
where the second equality is understood by writing Eq.~(\ref{foo00}) in bra-ket notation, as
\begin{equation}\label{foo19}
\ket{\dot{r}} = \ket{\partialbar_\gamma r} + {\cal S}\ket{\dot{\xv}} = 0\,.
\end{equation}
Notice that the stress tensor for athermal elastic materials is defined as $\sigma\!\equiv\! V^{-1}dU/d\gamma$ \cite{lutsko}, and therefore the form of Eq.~(\ref{foo10}) is expected; up to a factor of $V^{-1}$, it has the Irving-Kirkwood form of the edge forces' contribution to the stress tensor. Eq.~(\ref{foo09}) now becomes
\begin{equation}
dU/d\gamma = \partialbar U/\partialbar\gamma -\braket{\tau}{\partialbar_\gamma r}\,,
\end{equation}
which allows us to easily carry out another full derivative with respect to strain under the rigid-edge and mechanical-equilibrium constraints, as
\begin{eqnarray}
\!\!\!\! d^2U/d\gamma^2& = & \partialbar^2 U/\partialbar\gamma^2 + \braket{\partial_\xv \partialbar_\gamma U}{\dot{\xv}} \nonumber\\
&&- \braket{\tau}{\partialbar^2_{\gamma,\gamma}r} - \bra{\tau}\partial_\xv\partialbar_\gamma r\ket{\dot{\xv}} - \braket{\dot{\tau}}{\partialbar_\gamma r}\,, \label{foo11}
\end{eqnarray}
where we denote $\partialbar^2_{\gamma,\gamma}\=\partialbar^2/\partialbar\gamma^2$, and $\partial_\xv\partialbar_\gamma r$ is a linear operator such that
\[
\bra{\tau}\partial_\xv\partialbar_\gamma r\ket{\dot{\xv}} = \sum_{\mbox{\tiny edges }ij}\tau_{ij}\frac{\partial}{\partial\xv_\ell}\frac{\partialbar r_{ij}}{\partialbar\gamma}\cdot\dot{\xv}_\ell\,.
\]
Up to a factor of $V^{-1}$ (see Eq.~(\ref{foo12})), Eq.~(\ref{foo11}) constitutes an atomistic expression for the elastic moduli of our athermal elastically-embedded rigid-edge frames, that are uniquely determined by the nodes' coordinates.

\section{Numerical methods, models and protocols}
\label{numerics}
Having described in detail our theoretical framework, we next turn to concisely reviewing the numerical methods employed in our work, that we use to validate our micromechanical theory. A comprehensive description of our numerical methods, procedures and protocols are provided in Appendix~\ref{numerics_appendix}.  

\begin{figure}[t]
\centering
\includegraphics[width = 0.5\textwidth]{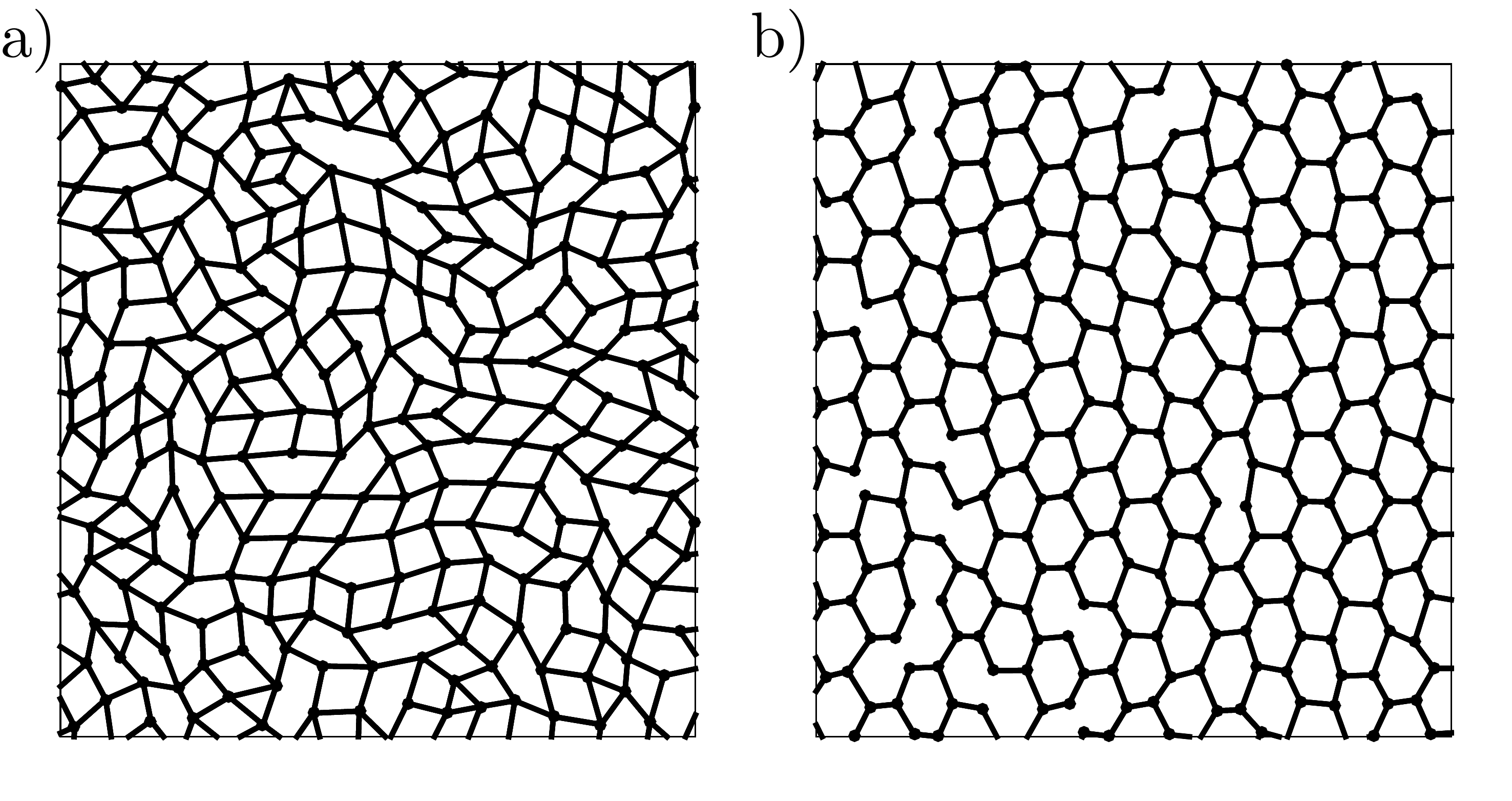}
\caption{\footnotesize Examples of the floppy frames employed for our study: (a) a packing-derived network obtained by diluting the edges of a packing's contact network, and (b) a diluted off-lattice honeycomb network with disorder of the node positions ($d_{\text{max}}\!=\!0.5$), see Appendix~\ref{numerics_appendix} for a detailed description of the protocol used to construct these frames.}
\label{exampleNetwork}
\end{figure}

To establish the generality of our results, we simulated two types of embedded frames: frames derived from packings of soft discs by inheriting and systematically pruning their contact networks, and frames obtained by perturbing the nodes of honeycomb lattices, and pruning their edges. Examples of these disordered frames are shown in Fig.~\ref{exampleNetwork}. 

To further reinforce the generality of our approach, we have also selected two forms for the embedding elastic energy in which our floppy frames are embedded. The first potential energy function depends quadratically on the angles formed between pairs of edges that share a common node, and is meant to mimic the bending energy of biopolymer fibers. The second potential energy is a simple Hookean-spring interaction that is introduced between nearby pairs of nodes that are not already connected by a rigid edge of the floppy frame. The precise functional form of the embedding potential energies and further details can be found in Appendix~\ref{numerics_appendix}. 

We have developed athermal, quasistatic deformation simulations derived from the formalism developed in Sect.~\ref{theoretical_framework}. In these simulations our embedded frames are deformed under simple shear or expansive strains, as described in Subsect.~\ref{node_dynamics}. This amounts to integrating the equations of motion (\ref{foo15}) and (\ref{foo16}) by iteratively calculating the nonaffine velocities $\ket{\dot{\xv}}$ and the deformation-induced variation of the edge forces $\ket{\dot{\tau}}$ using Eq.~(\ref{foo03}), applying small incremental strain steps, and evolving the configuration and edge forces accordingly. We further exploited our theoretical framework to apply correction steps that systematically eliminate --- to any chosen precision --- the accumulated integration errors that stem from employing finite integration steps, see complete derivation and description in Appendix~\ref{numerics_appendix}. During the deformation of the networks plastic instabilities can occur. A local rearrangement drives the network towards a different stable configuration. In Appendix~\ref{plastic_instabilities} we provide an example of such an instability and discuss their influence on the measured observables. 
  
\section{Micromechanical theory of strain stiffening and numerical validation}
\label{results}

\subsection{Elastic moduli in undeformed states}
\label{isotropic}

We kick off the discussion by considering the coordination dependence of elastic moduli of \emph{undeformed} (i.e.~to which no strain has yet been applied) elastically-embedded frames. 
Recall first our assumption that before any imposed deformation the edge forces $\ket{\tau}$ and the potential-derived forces $\ket{\partial_\xv U}$ both identically vanish. Consequently, it is convenient to consider the form for elastic moduli $E_{\gamma\!=\!0}$ of undeformed systems obtained by setting $\ket{\partial_\xv U}\=\zerovector$ in Eq.~(\ref{foo09B}), namely
\begin{equation}\label{foo21}
E_{\gamma\!=\!0} = \frac{\partialbar^2_{\gamma,\gamma}U + 2\braket{\partial_\xv\partialbar_\gamma U}{\dot{\xv}} + \bra{\dot{\xv}}{\cal M}\ket{\dot{\xv}} }{V}\,.
\end{equation}

We next 
utilize a mean-field approximation; we consider a simplified embedding elastic energy $U$ that consists of connecting each node to its \emph{absolute} initial position by a Hookean spring with unit stiffnesses. In this simplified case, following the definition of the operator ${\cal A}$ (see Eq.~(\ref{foo13}) and discussion in Subsect.~\ref{node_dynamics}) one finds ${\cal A}\={\cal M}\={\cal I}$, and recall importantly that $\ket{\tau}\=0$ in undeformed frames. This latter condition implies that $\braket{\partialbar_\gamma \fv}{\partialbar_\gamma \fv}$ is regular (see Eq.~(\ref{foo20})), and therefore within our mean-field approximation Eq.~(\ref{foo16}) can be written as
\begin{equation}
\ket{\dot{\xv}} \simeq -{\cal S}^T\big( {\cal S}{\cal S}^T \big)^{-1}\big( \ket{\partialbar_\gamma r}\!+\!{\cal S}\ket{\partialbar_\gamma\fv} \big)\,.
\end{equation}
The characteristic scale of nonaffine velocities squared reads
\begin{equation}\label{foo17}
\dot{x}^2 \equiv \braket{\dot{\xv}}{\dot{\xv}}/N \sim \bra{b}\big({\cal S}{\cal S}^T\big)^{-1}\ket{b}/N\,,
\end{equation}
where $\ket{b}\!\equiv\!\ket{\partialbar_\gamma r}\!+\!{\cal S}\ket{\partialbar_\gamma\fv}$ is a vector with regular components. 

\begin{figure}[t]
\centering
\includegraphics[width = 0.5\textwidth]{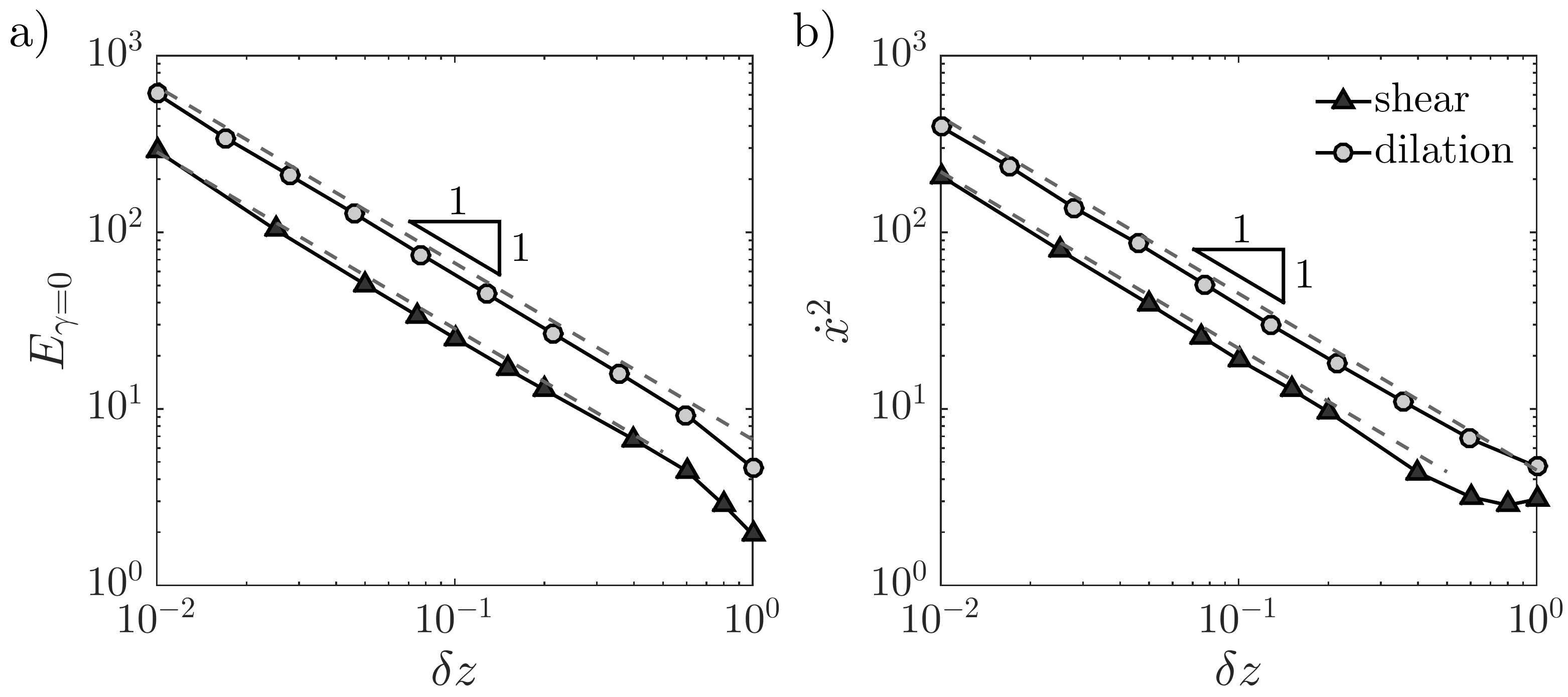}\\
\caption{\footnotesize For undeformed networks ($N = 12\,800$) of different coordinations $z$ we plot (a) the shear modulus $G$ (black triangles) and bulk modulus $K$ (grey circles), and (b) the characteristic scale of nonaffine velocities squared $\dot{x}^2\!\equiv\!\braket{\dot{\xv}}{\dot{\xv}}/N$ for both shear and dilation.}
\label{isotropic_mechanics}
\end{figure}

To proceed we introduce the spectral decomposition of the positive-definite operator ${\cal S}{\cal S}^T$ 
\begin{equation}\label{foo22}
{\cal S}{\cal S}^T = \sum_p \omega_p^2 \ket{\phi_p}\bra{\phi_p}\,,
\end{equation}
where the eigenvectors $\ket{\phi_p}$ and the squares of eigenfrequencies $\omega_p^2$ satisfy the eigenvalue equation ${\cal S}{\cal S}^T\ket{\phi_p}\=\omega_p^2\ket{\phi_p}$. The spectral properties of the operator ${\cal S}{\cal S}^T$ have been investigated extensively in \cite{asm_pnas, during2013phonon, asm_strain_stiffening}; in those works it has been shown that the distribution $D(\omega)$ of eigenfrequencies $\omega$ of the operator ${\cal S}{\cal S}^T$ in isotropic random floppy networks features a gap at low frequencies, and the emergence of a plateau of modes that follows $D(\omega)\!\sim\!\mbox{constant}$~above the characteristic frequency $\omega^*\!\sim\!\delta z\!\equiv\! z_c\!-\! z$ with $z_c\!=\!2\dbar$ denoting the Maxwell threshold in $\dbar$ dimensions. Examples of the spectra of ${\cal S}{\cal S}^T$, calculated for frames at two different coordinations, are shown in Fig.~\ref{omega_min_fig}a. These aformentioned details can be incorporated into Eq.~(\ref{foo17}) to obtain an estimation of the magnitude squared of the nonaffine velocities, as
\begin{equation}
\dot{x}^2 \sim \sum_p  \frac{\braket{b}{\phi_p}^2}{\omega_p^2} \sim \int_{\omega^*}^1\frac{D(\omega)}{\omega^2}d\omega \sim \frac{1}{\delta z}\,.
\end{equation}
Our analysis indicates that as $\delta z\!\to\!0$ the nonaffine velocities should diverge as $\dot{x}^2\!\sim\!1/\delta z$. In Fig.~\ref{isotropic_mechanics}b we plot $\dot{x}^2$ as measured in our floppy frames, against $\delta z$; we find perfect agreement with our mean-field prediction, supportinig that this scaling law (and others discussed below) is invariant to the particular functional form of the elastic energy $U$. In~\cite{wyart2008elasticity} the same scaling is predicted using a different line of argumentation.

Having established that the nonaffine velocities are singular as $\delta z\!\to\!0$, it becomes clear by examining Eq.~(\ref{foo21}) that to leading order 
\begin{equation}
E_{\gamma\!=\!0} \sim \dot{x}^2 \sim \frac{1}{\delta z}\,,
\end{equation}
in perfect agreement with our measurements shown in Fig.~\ref{isotropic_mechanics}a. 

\begin{figure}[h]
\centering
\includegraphics[width = 0.5\textwidth]{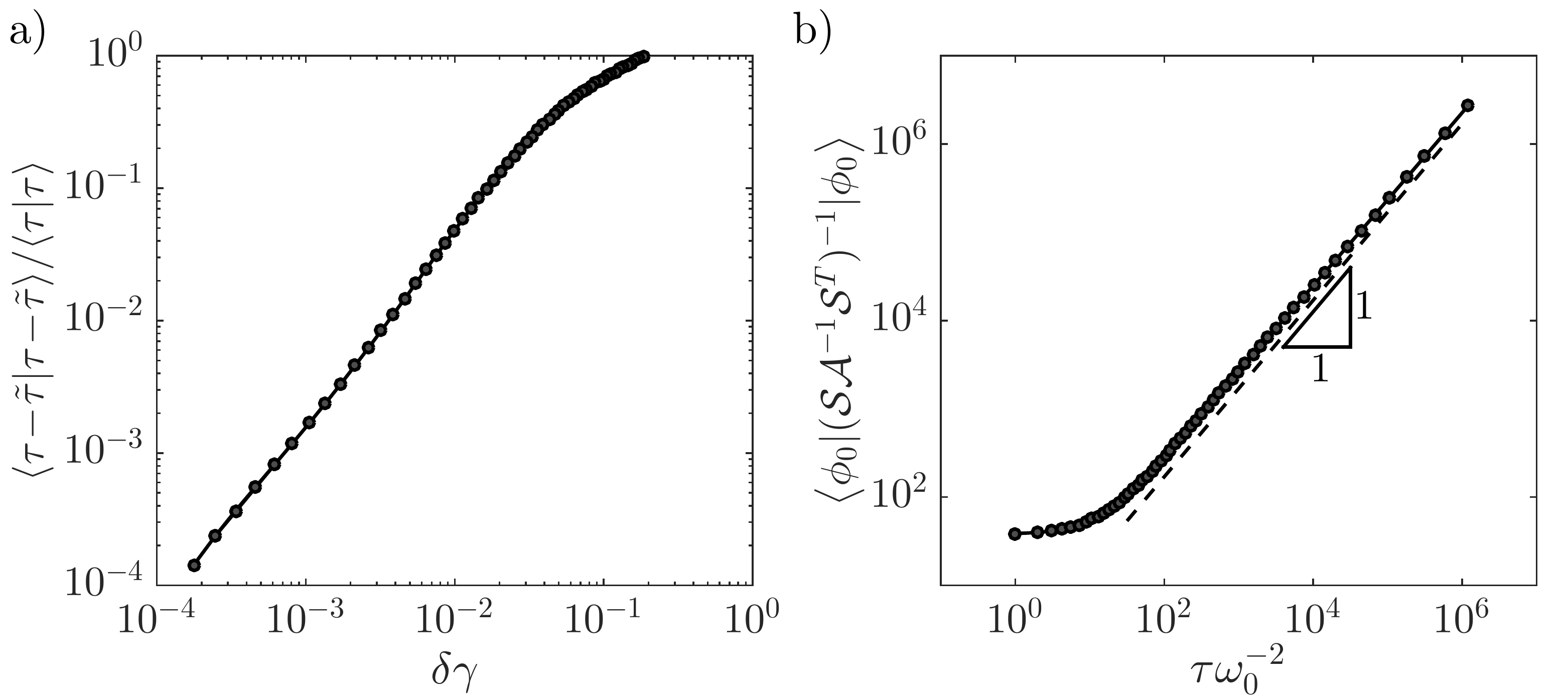}
\caption{ {\footnotesize For a honeycomb-lattice ($N\!=\!3\,600$, $z\!=\!3.0$, $d_{\text{max}}=0.5$),  (a) the approximation of the edge forces as given by Eq.~(\ref{foo34}) and denoted here by $\tilde \tau$ is tested by plotting the relative magnitude squared of their difference with the actual forces $\tau$ (see Eq.~(\ref{foo08})), as a function of the distance to the critical strain $\delta \gamma$. (b) Validation of relation (\ref{foo33}); for the same deformed network we plot $\bra{\phi_0}\big({\cal S}{\cal A}^{-1}{\cal S}^T\big)^{-1}\ket{\phi_0}$ against $\tau/\omega_0^2$, to find a linear relation.} }
\label{overlaps}
\end{figure}

\begin{figure*}[t]
\centering
\includegraphics[width = 0.5\textwidth]{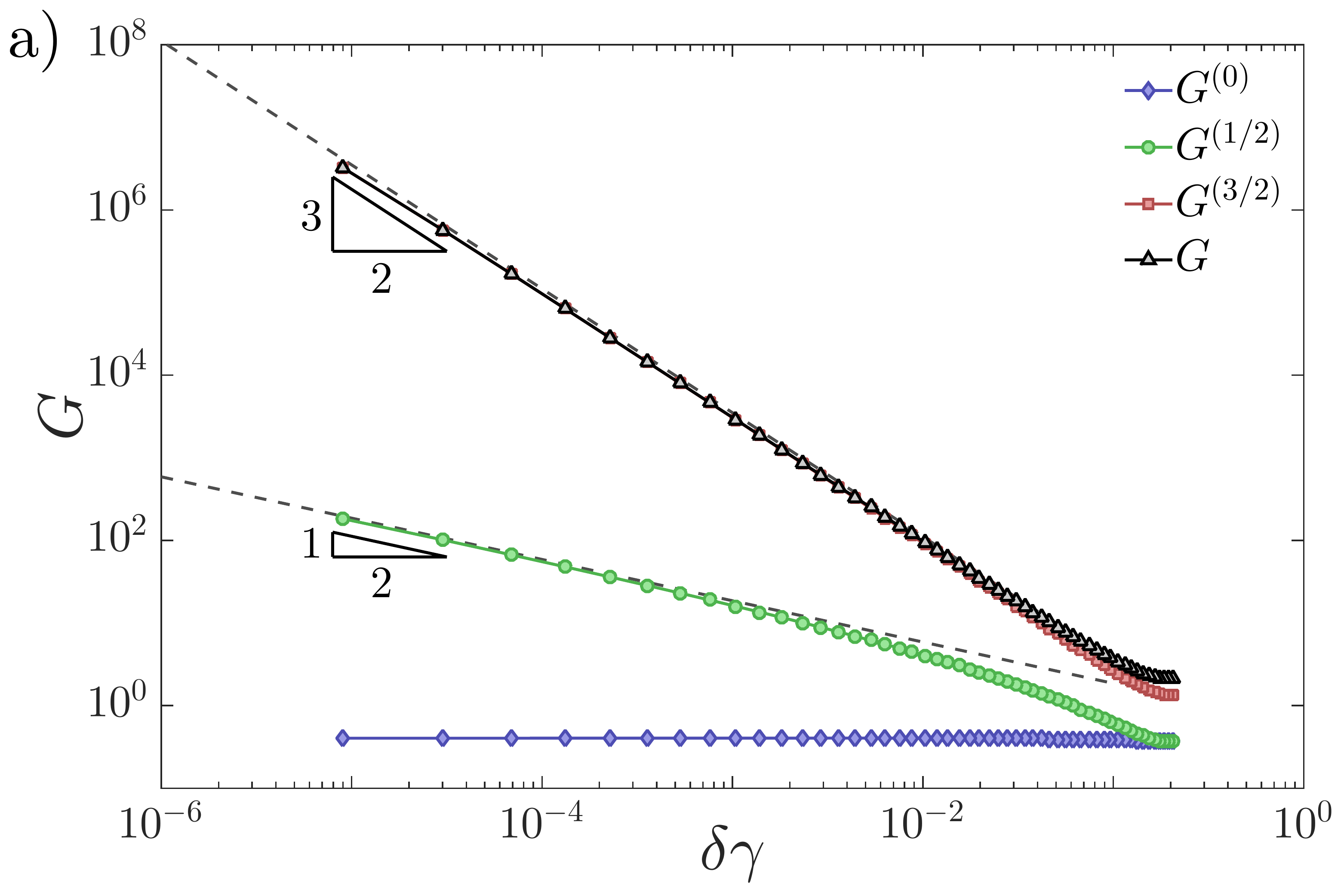}\includegraphics[width = 0.5\textwidth]{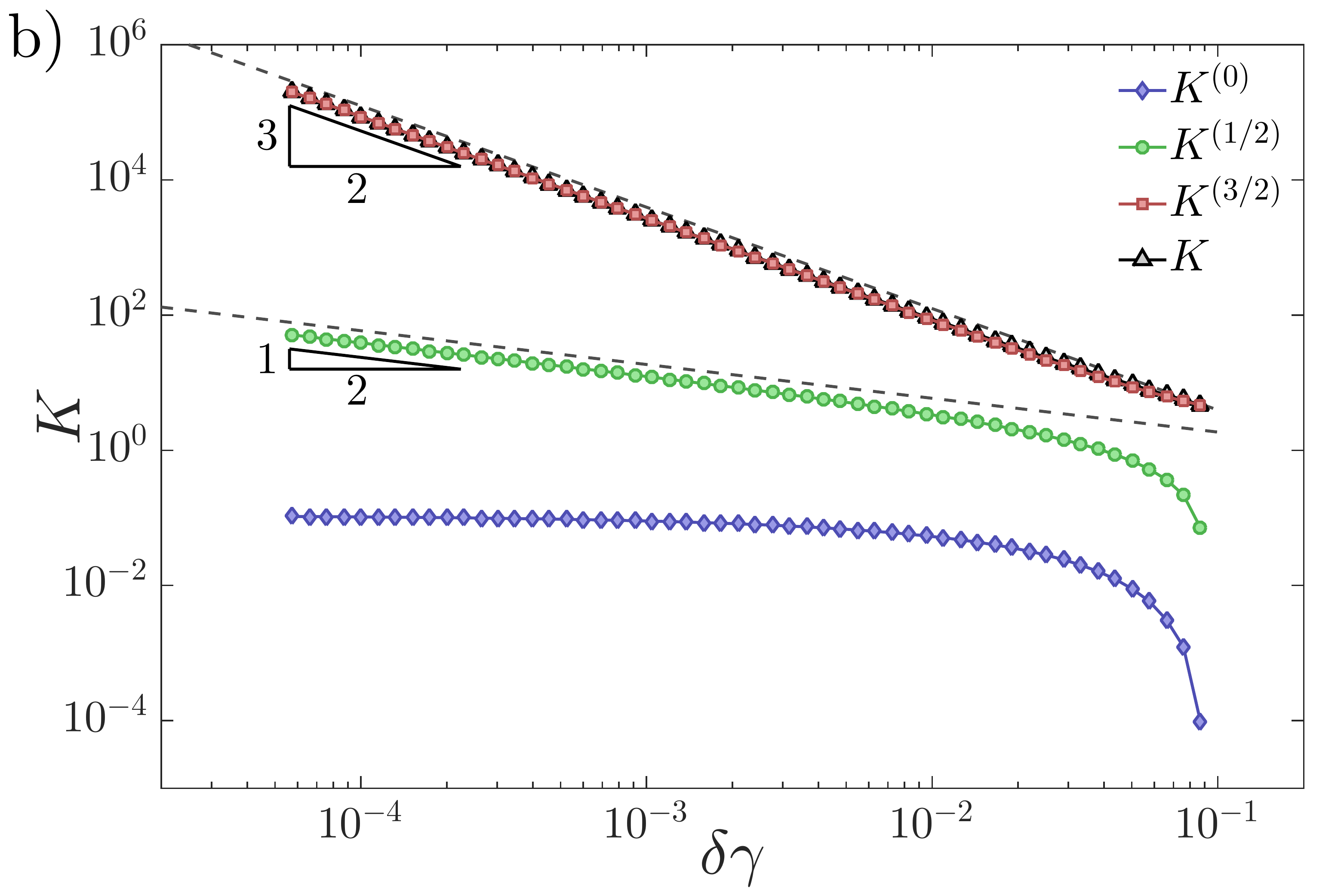}\caption{\footnotesize Elastic moduli of deformed, elastically-embedded floppy frames of rigid edges, as a function of the distance to the critical strain $\delta \gamma\!\equiv\!\gamma_c\!-\!\gamma$. Shown are the decomposition of the moduli's constituent terms, grouped by their scaling behavior with respect to strain, as expressed by Eqs.~(\ref{foo39})-(\ref{foo28}), and predicted by our theory. Panel (a) shows the shear modulus $G$ of a sheared disordered honeycomb-lattice ($N\!=\!3\,600$, $z\!=\!3.0$, $d_{\text{max}}\!=\!0.5$) with a bond-bending interactions as the embedding elastic energy, while panel (b) shows the bulk modulus $K$ of a packing-derived disordered frame with $z\!=\!3.0$ embedded in an Hookean-spring elastic network.}
\label{decomposition_of_modulus}
\end{figure*}

\subsection{Elastic moduli of strained configuration}

We now turn to our main focus -- the study of strain stiffening as manifested by the variation of elastic moduli upon deforming elastically-embeded floppy frames of rigid edges. We start by noting that at the limit of the strain stiffening transition the edge forces approach a self-balancing set, namely 
\begin{equation}
\lim_{\gamma\to\gamma_c}\frac{{\cal S}^T\ket{\tau}}{\sqrt{\braket{\tau}{\tau}}} = \zerovector\,.
\end{equation}
This implies that a zero mode of the operator ${\cal S}{\cal S}^T$ developes as $\gamma\!\to\!\gamma_c$, as indeed demonstrated in \cite{asm_strain_stiffening}. We denote this vanishing mode by $\ket{\phi_0}$ and its associated eigenvalue by $\omega_0^2$. Due to the vanishing of $\omega_0^2$ close to $\gamma_c$, the edge forces $\ket{\tau}$ as given by Eq.~(\ref{foo08}) can be approximated by
\begin{equation}
\ket{\tau} \simeq \frac{\bra{\phi_0}{\cal S}\ket{\partial_\xv U}}{\omega_0^2}\ket{\phi_0} = \frac{\braket{\Psi_0}{\partial_\xv U}}{\omega_0}\ket{\phi_0}\,,   \label{foo34} 
\end{equation}
where we assume that $\braket{\Psi_0}{\partial_\xv U}$ approaches a (coordination dependent) constant as $\gamma\!\to\!\gamma_c$, and notice crucially that ${\cal S}^T\ket{\phi_0}\!=\!\omega_0\ket{\Psi_0}$ \cite{asm_pnas}, where $\ket{\phi_0}$  and $\ket{\Psi_0}$ are normalized eigenvectors (see also discussion in Appendix~\ref{projection_operator_appendix}). The approximation Eq.~(\ref{foo34}) is validated numerically in Fig.~\ref{overlaps}a. In turn, the vanishing mode of ${\cal S}{\cal S}^T$ implies that the characteristic force in the edges
\begin{equation}
\label{fuu00}
\tau \equiv \sqrt{\braket{\tau}{\tau}/N} \sim \frac{1}{\omega_0}
\end{equation}
diverges as $\gamma\!\to\!\gamma_c$; we assume in what follows a power-law divergence $\tau\!\sim\!\delta\gamma^{-\chi}$, and aim at formulating a micromechanical derivation of the exponent $\chi$.

The divergence of the edge forces $\tau\!\sim\!\delta\gamma^{-\chi}$ leads to a stronger divergence of their deformation-induced variations $\dot{\tau}\!\sim\!\delta\gamma^{-(\chi+1)}$. Glancing at Eq.~(\ref{foo11}) for the elastic moduli, we conclude that the leading order term is the one involving $\ket{\dot{\tau}}$, and in particular we expect $E\!\sim\!\dot{\tau}$ where $\dot{\tau}\!\equiv\!\sqrt{\braket{\dot{\tau}}{\dot{\tau}}/N}$ is the characteristic scale of the deformation-induced variation of edge forces. 

The deformation-induced variations of the edge forces $\ket{\dot{\tau}}$ are spelled out in Eq.~(\ref{foo15}); to relate the latter to the characteristic edge force $\tau$ and the vanishing eigenvalue $\omega_0^2$, we assume that there are merely weak correlations between the eigenfunctions of the operators ${\cal A}$ and ${\cal S}{\cal S}^T$, allowing us to write a key scaling relation expected to be valid as $\gamma\to\gamma_c$, 
\begin{equation}\label{foo33}
-({\cal S}{\cal A}^{-1}{\cal S}^T)^{-1} \sim \tau({\cal S}{\cal S}^T)^{-1} \sim \tau\frac{\ket{\phi_0}\bra{\phi_0}}{\omega_0^2}\,,
\end{equation}
where the factor of $\tau$ can be understood by considering the definition of ${\cal A}$ as seen in Eq.~(\ref{foo13}). Relation (\ref{foo33}) is put to a direct numerical test in Fig.~\ref{overlaps}b, where we plot $\bra{\phi_0}\big({\cal S}{\cal A}^{-1}{\cal S}^T\big)^{-1}\ket{\phi_0}$ vs.~$\tau/\omega_0^2$ and find a linear relation between the two, establishing the validity of the aformentioned assumption of weak correlations between the eigenfunctions of the operators ${\cal A}$ and ${\cal S}{\cal S}^T$. 

Using the approximation Eq.~(\ref{foo33}) in Eq.~(\ref{foo15}), the edge force variations can be written as
\begin{equation}\label{foo41}
\ket{\dot{\tau}} \simeq \tau\frac{\braket{\phi_0}{\partialbar_\gamma r}}{\omega_0^2}\ket{\phi_0}\,,
\end{equation}
since the term involving $\ket{\partialbar_\gamma\fv}$ in $\ket{\dot{\tau}}$ is subdominant close to the critical strain $\gamma_c$, as argued in Appendix~\ref{remainbound}. We thus expect $\dot{\tau}\!\sim\!\tau/\omega_0^2$, and together with Eq.~(\ref{fuu00}) we obtain 
\begin{equation}
\dot{\tau}\sim \tau^2 \,\, \Rightarrow \,\, \chi=  1/2\,,
\end{equation}
leading to the conclusions
\begin{equation}
\tau \sim \frac{1}{(\gamma_c - \gamma)^{1/2}}\quad\mbox{and}\quad \dot{\tau} \sim \frac{1}{(\gamma_c - \gamma)^{3/2}}\,.
\end{equation}
In addition, from relation (\ref{foo10}) we conclude that $\dot{x}\!\sim\!\tau$, hence the characteristic nonaffine velocities should diverge as 
\begin{equation}
\dot{x}\sim\frac{1}{(\gamma_c-\gamma)^{1/2}}\,.
\end{equation}

To test our theoretical predictions, we have deformed our elastically-embedded frames using the methods described in Sect.~\ref{numerics}, and measured their elastic moduli using the microscopic expressions derived in Sect.~\ref{elastic_moduli}. In Fig.~\ref{decomposition_of_modulus}a we report the deformation-induced stiffening of the shear modulus in a single realization of an embedded floppy frame with $z\=3.0$. The triangular symbols represent the full shear modulus as given by Eq.~(\ref{foo11}), which is plotted against the strain difference $\delta \gamma\!\equiv\!\gamma_c\!-\!\gamma$ to the critical stiffening strain $\gamma_c$. As our key result, we find a very clean $G \sim \delta \gamma^{-3/2}$ scaling over several orders of magnitude of $\delta \gamma$. In Fig.~\ref{decomposition_of_modulus}b a similar behavior is observed for the bulk modulus  in networks under expansion.

We also plot in Fig.~\ref{decomposition_of_modulus} the various contributions to the elastic moduli $E$; to this aim we define
\begin{equation}
E = E^{(0)} + E^{(1/2)} + E^{(3/2)}\,,
\end{equation}
where, following Eq.~(\ref{foo11}) 
\begin{eqnarray}
E^{(0)} & = & \frac{\partialbar^2 U/\partialbar\gamma^2}{V}\,, \label{foo39} \\
E^{(1/2)} & = &\frac{\braket{\partial_\xv\partialbar_\gamma U}{\dot{\xv}} - \braket{\tau}{\partialbar^2_{\gamma,\gamma}r} - \bra{\tau}\partial_\xv\partialbar_\gamma r\ket{\dot{\xv}}}{V}\,, \label{foo38}  \\
E^{(3/2)} & = & -\frac{\braket{\dot{\tau}}{\partialbar_\gamma r}}{V}\,. \label{foo28}
\end{eqnarray}
Our data indicates that $E^{(0)}\!\sim\!\delta\gamma^{0}$, $E^{(1/2)}\!\sim\!\delta\gamma^{-1/2}$, and $E^{(3/2)}\!\sim\!\delta\gamma^{-3/2}$, in perfect agreement with the scaling relations derived above. In Appendix~\ref{robbiez_contraction} we explain why $\bra{\tau}\partial_\xv\partialbar_\gamma r\ket{\dot{\xv}}\!\sim\!\delta\gamma^{-1/2}$ despite that $\tau\!\sim\!\dot{x}\!\sim\!\delta\gamma^{-1/2}$.

\begin{figure}[t]
\centering
\includegraphics[width = 0.5\textwidth]{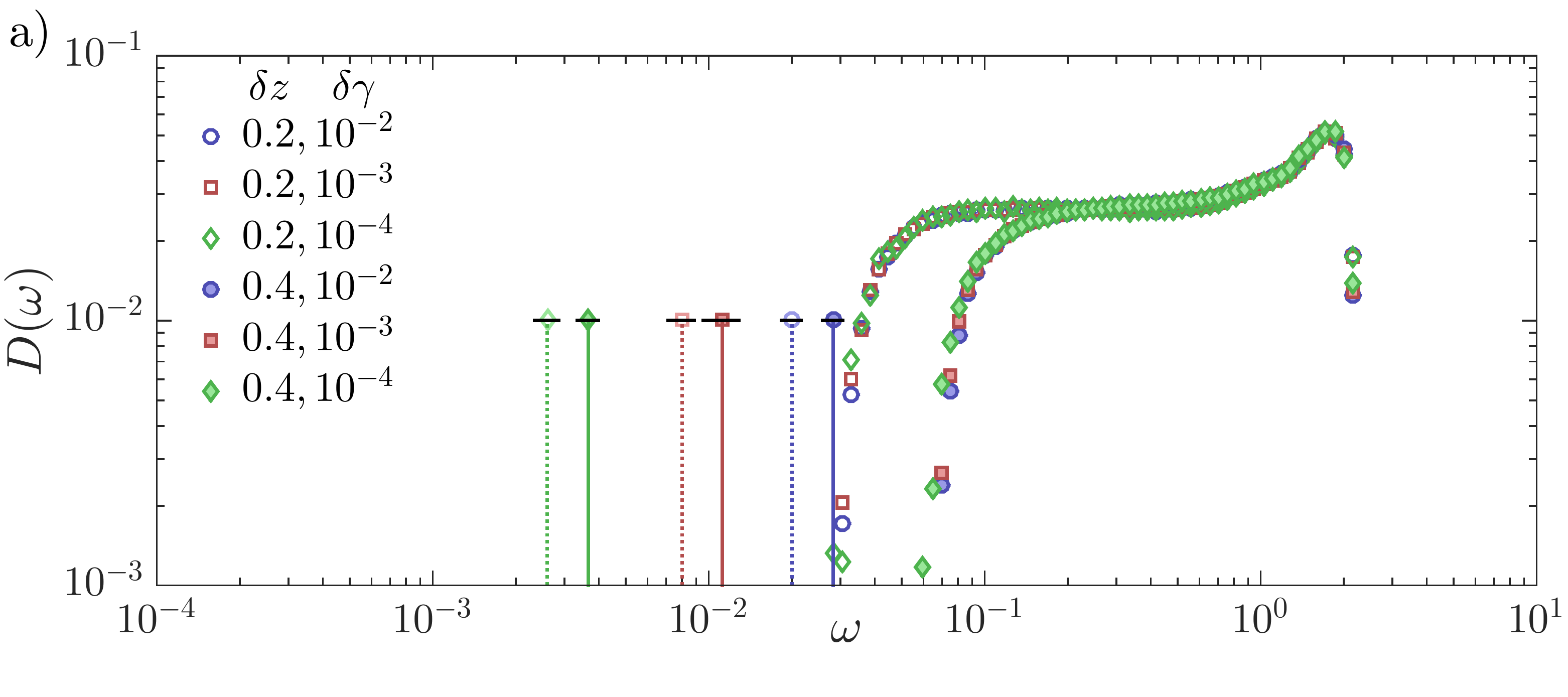}
\includegraphics[width = 0.5\textwidth]{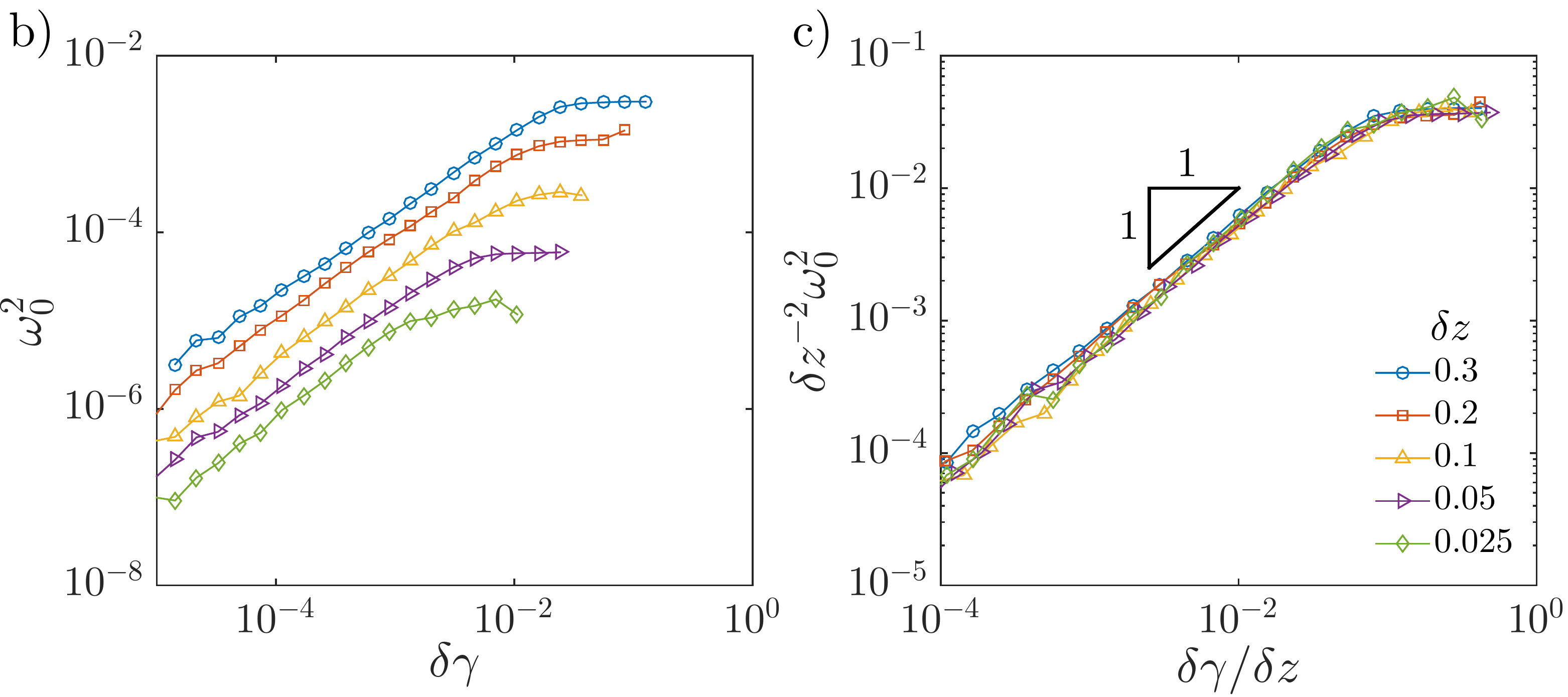}
\caption{\footnotesize  (a) The distribution of eigenfrequencies $D(\omega)$ of the operator ${\cal S}{\cal S}^T$, calculated for sheared networks of $N\!=\!4\,096$ nodes with coordinations $z=3.6$ and $z=3.8$, at different stages of the deformation. Each curve shows the distribution of eigenfrequencies binned over 100 independent realizations. The distributions feature a plateau above an onset frequency $\omega^*\!\sim\!\delta z\!\equiv\! z_c\!-\! z$, with $z_c$ the Maxwell threshold. Upon deformation, a single, lowest frequency mode $\omega_0$ per realization escapes from the plateau, and vanishes as $\omega_0\!\sim\!\sqrt{\delta\gamma}$ upon approaching the critical strain, as shown in panels (b),(c). The mean (over realizations) frequency of the lowest mode $\omega_0$ is indicated with a vertical line, and its standard deviation is indicated by horizontal lines. (b) The lowest eigenvalue $\omega_0^2$ of the operator ${\cal S}{\cal S}^T$ as function of the distance to the critical point $\gamma_c$ for packing derived networks ($N = 1\,600$) of different coordination number $\delta z$. Each data-point represents the median over 20 realization. (c) The same data as presented in panel (b), recasted into the scaling form given by Eq.~(\ref{foo36}). }
\label{omega_min_fig}
\end{figure}

The predicted scaling laws $\omega_0^2\!\sim\!\delta\gamma$ and $E\!\sim\!\dot{\tau}\!\sim\!\delta\gamma^{-3/2}$ do not capture the possible coordination dependence of these observables. To resolve the coordination dependence of the vanishing eigenvalue $\omega_0^2$ and of the elastic modulus $E$, we first note that strain stiffening sets in at a characteristic strain scale $\delta\gamma_\star\!\sim\!\delta z$ \cite{wyart2008elasticity,asm_strain_stiffening}. At strains $\gamma_c\!-\!\gamma \!\lesssim\! \delta\gamma_\star$ we expect $\omega_0^2\!\sim\!\delta\gamma$ as derived above. On the other hand, in isotropic, undeformed states, one expects $\omega_0^2\!\sim\!\delta z^2$, as shown e.g.~in \cite{during2013phonon}. We therefore write a scaling ansatz for the vanishing eigenvalue $\omega_0^2$ of the form
\begin{equation}\label{foo36}
\omega_0^2 \sim \delta z^2 {\cal F}_1\left(\frac{\delta\gamma}{\delta z}\right)\,,
\end{equation}
where the scaling function ${\cal F}_1(x)\!\sim\!x$ for $x\!\ll\!1$, and ${\cal F}_1(x)\!\sim\!\mbox{constant }$ for $x\!\gg\!1$.

In Fig.~\ref{omega_min_fig}b we plot the vanishing eigenvalue $\omega_0^2$ (defined as the minimal eigenvalue of ${\cal S}{\cal S}^T$) vs.~the strain difference to the stiffening transition $\delta\gamma$, for systems with various coordinations $z$ as indicated by the legend. The excellent agreement of our data with the scaling form Eq.~(\ref{foo36}) implies that the vanishing eigenmode depends on strain and coordination near the strain stiffening transition as
\begin{equation}
\omega_0^2 \sim \delta z\delta\gamma\,.
\end{equation}
This is one of the key results of our work. 
\begin{figure}[t]
\centering
\includegraphics[width = 0.5\textwidth]{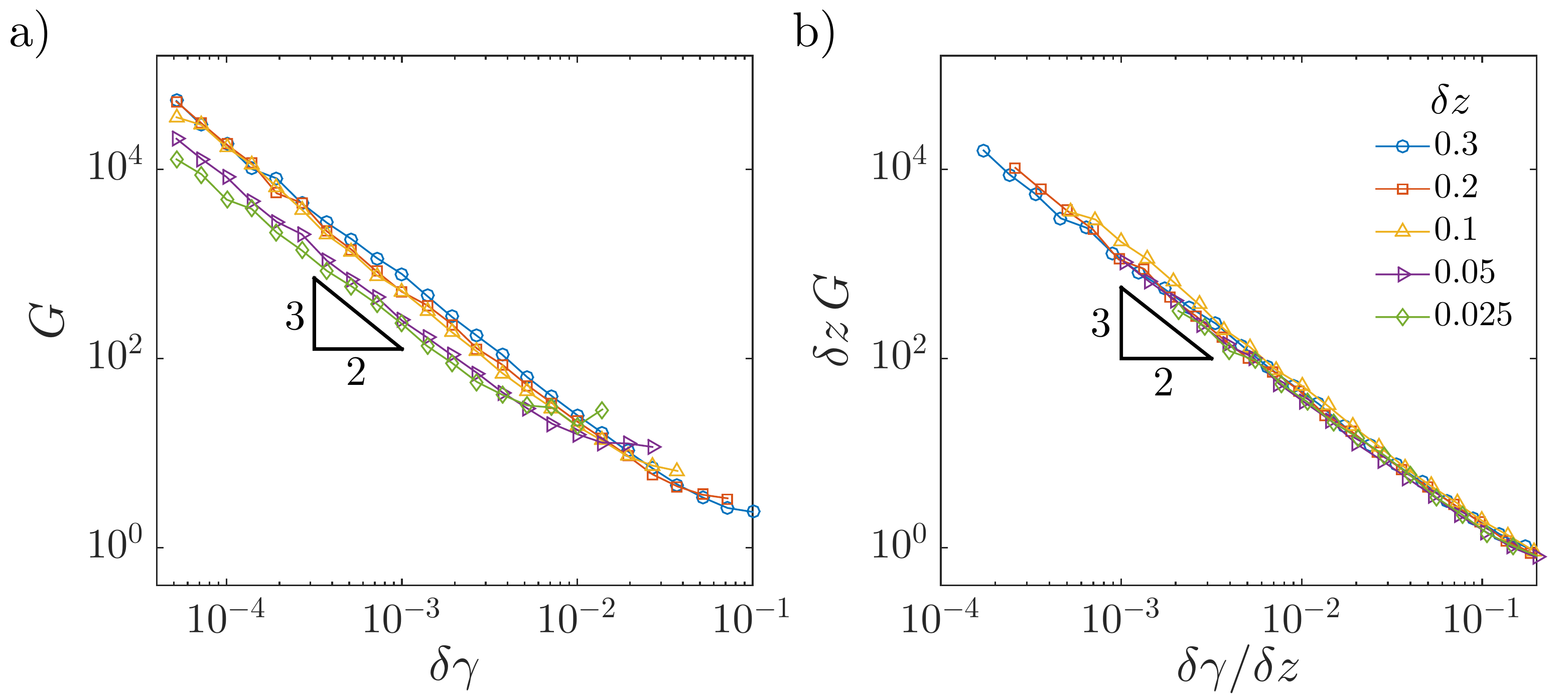}\\
\includegraphics[width = 0.5\textwidth]{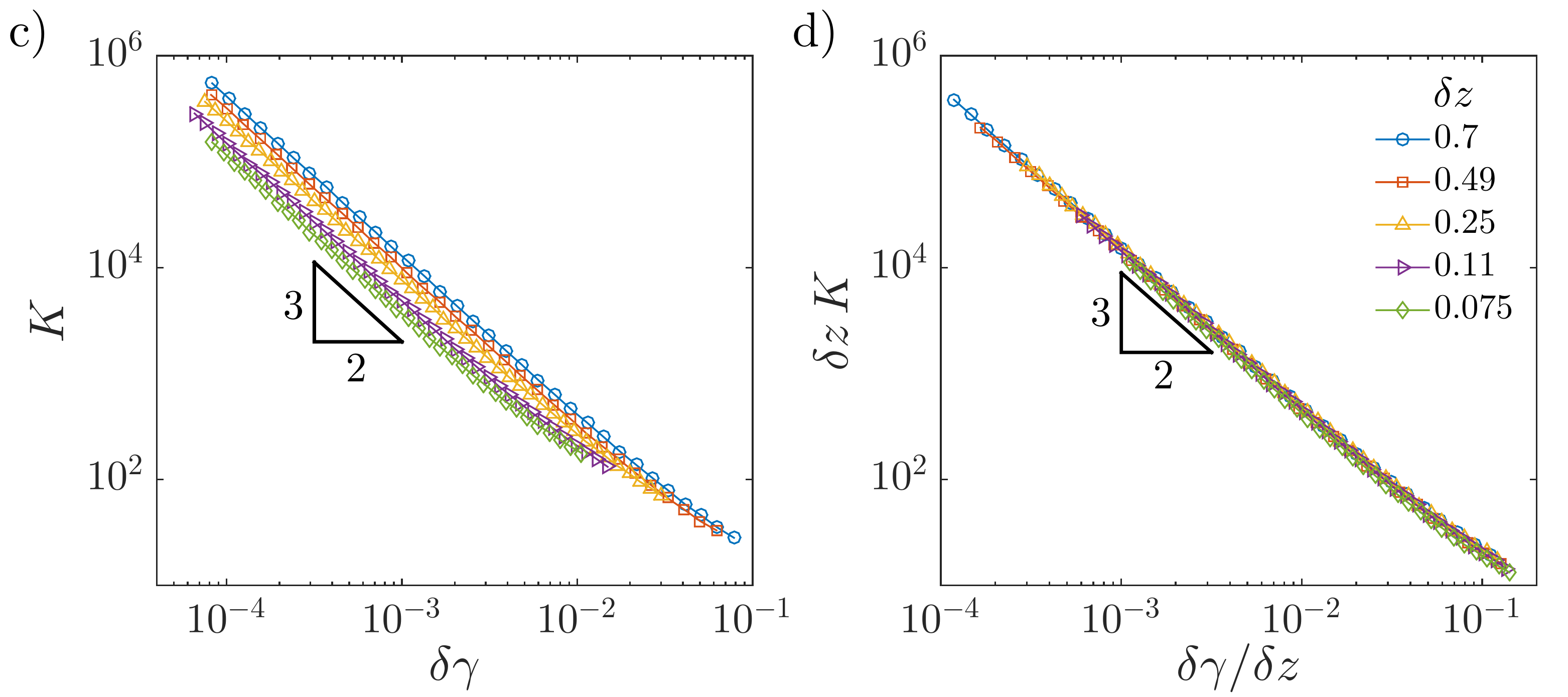}
\caption{\footnotesize (a) The shear modulus $G$ as function of the distance to the critical point $\gamma_c$ for packing derived networks of different coordination ($N = 1\,600$) number $\delta z$. Each curve is the result of the median of 20 realisation. (b) The shear modulus $G$ rescaled by $\delta z$ to obtain a collapse of the same data. (c) and  (d) are the same as (a) and (b) but for the bulk modulus $K$ of networks under expansive deformation.}
\label{moduli_fig}
\end{figure}

We finally turn to the coordination dependence of the elastic modulus. In Subsect.~\ref{isotropic} we have shown that in isotropic, undeformed states the elastic moduli scales as $\delta z^{-1}$, whereas in this Section we find $E\!\sim\!\delta\gamma^{-3/2}$ in deformed states approaching the strain stiffening transition. We combine once again these results together with the strain scale $\delta\gamma_\star\!\sim\!\delta z$ into a scaling ansatz
\begin{equation}\label{foo37}
E \sim \delta z^{-1}{\cal F}_2\left(\frac{\delta\gamma}{\delta z}\right)\,,
\end{equation}
where the scaling function ${\cal F}_2(x)\!\sim\!x^{-3/2}$ for $x\!\ll\!1$, and ${\cal F}_2(x)\!\sim$ constant for $x\!\gg\!1$.

In Fig.~\ref{moduli_fig}a we plot the shear modulus $G$ vs.~the strain difference to the stiffening transition $\delta\gamma$, for systems with various coordinations $z$ as indicated by the legend. The excellent agreement of our data with the scaling form Eq.~(\ref{foo37}) implies that elastic moduli depend on strain and coordination near the strain stiffening transition as
\begin{equation}
E \sim \sqrt{\delta z}\, \delta \gamma^{-3/2}\,.
\end{equation}
This is another key result of our work.


\subsection{Lower bound for the critical exponent $\chi$}

The main assumption made to derive the critical exponent $\chi$ regards the weak correlations between the matrices $\cal A$ and ${\cal S}$. This assumption can not always be guaranteed, for example for a freely-jointed chain, which correspond to the limiting case of a coordination two network, the weak correlations assumptions seems to be false. Notwithstanding, in such cases we can still establish a lower bound for the critical exponent $\chi$, which is saturated in the cases studied in this work, when $\cal A$ and ${\cal S}$ are weakly correlated. 
 
We denote by $\alpha_{\text{min}}$ the smallest eigenvalue of the matrix ${\cal S}{\cal A}^{-1}{\cal S}^T$. Given any arbitrary vector $\ket{c}$, one has
\begin{eqnarray}
\label{bound1}
\frac{\bra{c}\big({\cal S}{\cal A}^{-1}{\cal S}^T\big)^{-2}\ket{c}}{\braket{c}{c}}\leq \frac{1}{\alpha^2_{\text{min}}},\\
\label{bound2}
\frac{\bra{c}{\cal S}{\cal A}^{-1}{\cal S}^T\ket{c}}{\braket{c}{c}}\geq \alpha_{\text{min}}.
\end{eqnarray}
In addition we can rewrite 
\begin{equation} 
\frac{\bra{c}{\cal S}{\cal A}^{-1}{\cal S}^T\ket{c}}{\braket{c}{c}}=\frac{\bra{d}{\cal A}^{-1}\ket{d}}{\braket{d}{d}} \frac{\bra{c}{\cal S}{\cal S}^T\ket{c}}{\braket{c}{c}}\geq \frac{\omega_0^2}{\lambda_{\text{max}}}.
\label{bound3}
\end{equation}
where $\ket{d}\!\equiv\!{\cal S}^T\ket{c}$ and $\lambda_{\text{max}}$ denotes the largest eigenvalue of $\cal A$. The inequality (\ref{bound2}) saturates when $\ket{c}$ is set to be the lowest eigenvector of ${\cal S}{\cal A}^{-1}{\cal S}^T$. Now, since the upper bounds (\ref{bound2}) and (\ref{bound3}) are independent of the choice of $\ket{c}$, this implies that $\omega_0^2/\lambda_{\text{max}}\!\leq\!\alpha_{\text{min}}$. Finally, using the bound (\ref{bound1}) one finds 
 \begin{eqnarray}
 \label{fuu02}
  \frac{\bra{b}\big({\cal S}{\cal A}^{-1}{\cal S}^T\big)^{-2}\ket{b}}{\braket{b}{b}}\leq \frac{\lambda^2_{\text{max}}}{\omega_0^4}\,.
 \end{eqnarray}
The source of the singularity of $\cal A$ is the contraction with the edge forces $\ket{\tau}$ (see Eq.~(\ref{foo13})), thus $\lambda_{\text{max}}/\tau$ must remain bounded at the critical strain. Then there must exist a constant $B$ such that $\lambda_{\text{max}}\leq B \tau$. Combining (\ref{fuu00}),(\ref{foo33}) and (\ref{fuu02}) one finds
 \begin{equation}
 \dot{\tau}^2\sim \frac{\braket{\dot{\tau}}{\dot{\tau}} }{\braket{b}{b}}\leq \frac{B^2 \tau^2}{\omega_0^4}\sim \tau^6,
 \end{equation}
 which finally implies that 
 \begin{equation}
 -2(\chi+1)\geq -6\chi \, \Rightarrow \, \chi\geq  1/2
 \end{equation}
Remarkably, for any strain stiffening transition, independently of the elastic matrix dimension and topology of the embedded network, we expect a diverging elastic modulus $E\!\sim\!(\gamma_c-\gamma)^{-(\chi+1)}$ with a critical exponent $\chi\!\ge\!1/2$.

\subsection{Diverging lengthscale}
We end this Section with revealing the existence of an underlying diverging lengthscale that accompanies the critical strain-stiffening transition. The jamming literature offers numerous discussions and numerical investigations of diverging lengths close to various jamming transitions. A brief but rather complete and recent review of those previous efforts can be found in \cite{states_of_self_stress_epje}. 

\begin{figure}[t]
\centering
\includegraphics[width = 0.5\textwidth]{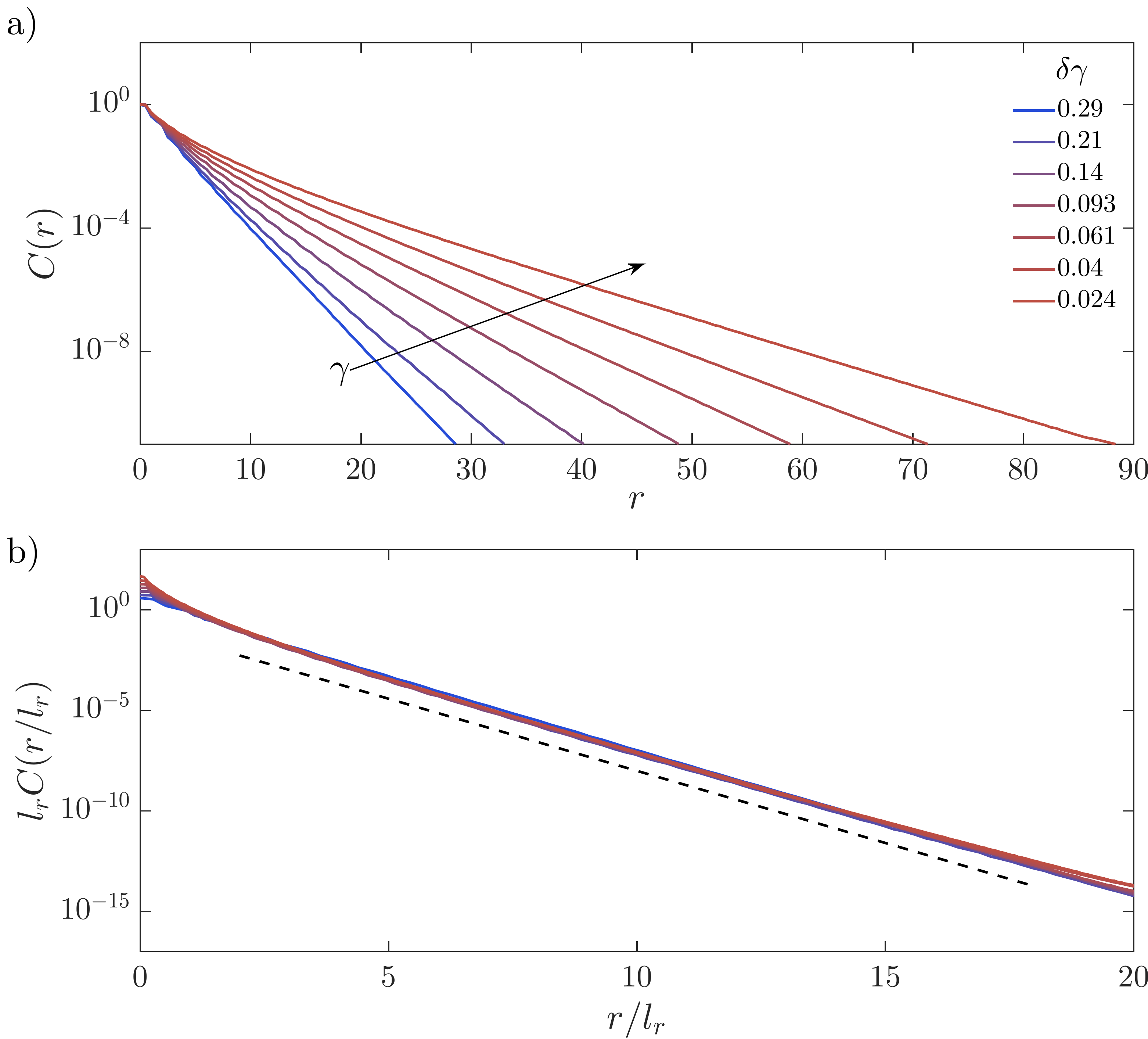}
\caption{\footnotesize (a) Decaying correlations $C(r)$ as function distance between pairs for a single network realisation $(N = 40\,000)$ under increasing amount of deformation reveals the increasing length scale in the system. (b) In the lower panel we present the collapse of the correlation functions by the length scale $l_r\!\sim\!1/\sqrt{\delta\gamma}$. }
\label{length_scale_fig}
\end{figure}

We find that the clearest observation of the diverging length is made as follows; we begin with considering the explicit form of the edge forces as given by Eq.~(\ref{foo08}). The latter can be decomposed as
\begin{equation}\label{foo32}
\ket{\tau} = ({\cal S}{\cal S}^T)^{-1}{\cal S}\ket{\partial_\xv U} = \sum_\alpha c_\alpha \ket{\tau_\alpha}\,,
\end{equation}
where $c_\alpha\!\equiv\!\bra{\alpha}{\cal S}\ket{\partial_\xv U}$ is the projection of the potential-derived forces onto the $\alpha\th$ edge, $\ket{\alpha}$ is an edge-wise vector which has zeros in all component besides the $\alpha\th$ one, and
\begin{equation}
\ket{\tau_\alpha} \equiv ({\cal S}{\cal S}^T)^{-1}\ket{\alpha}  \,.
\label{response1}
\end{equation}
Properties of the edge forces can therefore be determined by knowledge of the spectral properties of ${\cal S}{\cal S}^T$. It has been well-established that disordered floppy networks feature a frequency scale $\omega^*\!\sim\!\delta z$ above which there is a plateau of modes that remain statistically invariant under applied strain \cite{asm_strain_stiffening}, as also shown in Fig.~\ref{omega_min_fig}. In addition, at strains $\delta \gamma\!<\!\delta \gamma_\star$ the frequency scale $\omega_0$ appears below $\omega^*$, followed by a set of modes which in the thermodynamic limit should fill the gap between $\omega_0$ and $\omega^*$. Such modes can be considered as plane wave modulations of the vanishing mode $\ket{\phi_0}$ \cite{asm_strain_stiffening}. Then, from Eq.~(\ref{response1}) and using the spectral decomposition of ${\cal S}{\cal S}^T$ the component of $\ket{\tau_\alpha}$ pertaining to the $\beta\th$ edge reads
\begin{equation}
\braket{\beta}{\tau_\alpha}=\sum_{\omega_0\leq\omega <\omega^*}\frac{ \braket{\beta}{\phi_\omega}\braket{\phi_\omega}{\alpha}}{\omega^2} +\sum_{\omega^*\leq\omega }\frac{ \braket{\beta}{\phi_\omega}\braket{\phi_\omega}{\alpha}}{\omega^2} \,.
\label{response2}
\end{equation}
The second sum on the RHS of Eq.~(\ref{response2}) has been shown in \cite{during2013phonon} to feature an exponential decay $\sim e^{-r \sqrt{\delta z} }$ with $r$ the distance between the $\alpha\th$ and $\beta\th$ edges. The sum between $\omega_0$ and $\omega^*$ in Eq.~(\ref{response2}) has been shown \cite{asm_strain_stiffening} to also follow an exponential decay $\sim e^{-r/l_r}$ with $l_r\sim 1/\omega_0$. Therefore, close to the critical strain the second sum is subdominant, and $\braket{\beta}{\tau_\alpha}\!\sim\! e^{-r/l_r}$ with $l_r\sim 1/\sqrt{\delta \gamma}$. Our prediction is verified in Fig.~\ref{length_scale_fig}, where we show the average spatial decay of the squares $C(r) \!\equiv \!\overline{\braket{\beta}{\tau_\alpha}^2}$ as a function of the distance $r$ between the $\alpha\th$ and $\beta\th$ edges.

\section{Summary and outlook}
\label{conclusions}

In this work we have revealed the critical behavior that underlies the strain stiffening transition observed in athermal biopolymer networks. This transition has been traditionally probed using numerical models by introducing two stiffness scales characterizing bending and stretching modes respectively, and choosing the ratio $\tilde{\mu}$ between these stiffness scales to be very large in order to expose the underlying critical behavior \cite{PhysRevE.94.042407,broedersz2011criticality}, as demonstrated in Fig.~\ref{model_motivation_fig}. Here we \emph{directly} take the limit $\tilde{\mu}\!\to\! \infty$, and present the formalism and simulation method that emerges from this limit. 

The formalism introduced allows us to construct a scaling theory, both for undeformed, isotropic complex solids of floppy frames embedded in an elastic medium, and of the mechanics of such solids subjected to large deformations. Our main results are $(i)$ that undeformed, isotropic elastic moduli depend on coordination as $E\!\sim\!\delta z^{-1}$, with $\delta z\!\equiv\! z_c\! -\! z$ denoting the coordination difference to the Maxwell threshold $z_c\!\equiv\!2\dbar$ in $\dbar$ dimensions. $(ii)$ Elastic moduli of complex solids approaching the strain stiffening transition scale as $E\!\sim\!\sqrt{\delta z}\,\delta\gamma^{-3/2}$, with $\delta\gamma\!\equiv\!\gamma_c\!-\!\gamma$ denoting the strain difference to the critical stiffening strain $\gamma_c$. $(iii)$ A diverging length $l_r\!\sim\!\delta\gamma^{-1/2}$ accompanies the critical strain-stiffening transition. 

In related research efforts the strain stiffening transition has been analyzed in the framework of phase transitions \cite{sharma2016strain,fred_2018}; the stiffening transition was shown in these works to be captured by the scaling form
\begin{equation}
E \sim \mu |\delta \gamma|^f {\cal G}_{\pm}\left(\frac{\tilde \kappa}{|\delta \gamma|^\phi}\right) \,, 
\end{equation}
with critical exponents $f$ and $\phi$ and $\tilde \kappa \equiv \frac{\kappa}{l_0^2 \mu}$ . Our approach is only able to capture the stiffening below the critical point corresponding to  ${\cal G}_{-}( \tilde \kappa |\delta \gamma|^{-\phi})$. The scaling function ${\cal G}_{-}(x)\!\sim\! x $ , for  $x\!\ll\!1$, which implies that $E\!\sim\!\mu \tilde{\kappa}| \delta \gamma |^{f-\phi}$. To this extend we can conclude that the two critical exponents $f$ and $\phi$ are related by the constraint ${f-\phi} = -3/2$. The reported values of ${f-\phi}$ in the literature \cite{sharma2016strain} vary from $-1.34$ to $-1.77$ and in good agreement with the given constraint. We further note that in \cite{fred_2018} the nonaffine velocities were reported to follow $\dot x^2\!\sim\!\delta \gamma^{-3/2}$ for networks build with a different protocol, whereas our scaling theory predicts a scaling of $\dot x ^2\!\sim\!\delta \gamma^{-1}$.

Our analysis reveals that there exists a set of observables whose scaling with respect to the distance to the strain stiffening transition $\delta \gamma$ can be directly interchanged with the difference between their coordination and the Maxwell threshold $\delta z$. For instance, upon shearing our elastically-embedded floppy frames, the nonaffine velocities scale as $\dot{x}\!\sim\!\delta \gamma^{-1/2}$, whereas the nonaffine velocities of isotropic, undeformed frames scales with coordination as $\dot{x}\!\sim\!\delta z^{-1/2}$. This interchangeability indicates that there is at least a partial underlying equivalence between constraining the space of floppy modes by imposing (macroscopic) external deformation, and constraining it by increasing the (microscopic) connectivity of the floppy network. Understanding this connection calls for further investigation.

\acknowledgements
G.D. ~acknowledges funding from Millennium Science Initiative of the Ministry of Economy, Development and Tourism, grant ``Nuclei for Smart Soft Mechanical Metamaterials".
G.D.~acknowledges support from FONDECYT Grant No.~1150463.  E.L.~ acknowledges support from the Netherlands Organisation for Scientific Research (NWO) (Vidi grant no.~680-47-554/3259). 
R.R. and  E.L.~acknowledges support from the Delta Institute for Theoretical Physics (D-ITP consortium), a program of NWO that is funded by the Dutch Ministry of Education, Culture and Science (OCW).

\appendix

\section{Model of networks with separated bending and stretching energy scales}
\label{energeticmodel}
Fig.~(\ref{model_motivation_fig}) shows the strain-stiffening transition as seen our model of floppy frames of \emph{rigid} edges embedded in an elastic energy, together with the transition as seen in the convetionally-employed model of spring networks that feature different bending and stretching stiffnesses. In this Appendix we provide details about the convetional spring network model, employed e.g.~in \cite{rens2016nonlinear}. 

The energy of the system consists of two contributions: a pairwise spring interaction between connected nodes, and a bending interaction between pairs of bonds. The total energy $U$ is given by  
\begin{equation}\label{foo40}
U = \frac{\mu}{2} \sum_{\mbox{\tiny edges }i,j} \Delta r_{ij}^2 + \frac{\kappa}{2} \sum_{\mbox{\tiny triples }i,j,k} \Delta \theta_{ijk}^2 \,,
\end{equation}
where $\mu$ and $\kappa$ are stretching and bending stiffnesses that set the strength of the two types of interaction. The behavior across the strain-stiffening transition is controlled by the dimensionless number $\tilde{\mu} \equiv \frac{\mu l_0^2}{\kappa} $, which is typically set to be much larger than unity, corresponding to well-separated energy scales.

In this case, mechanical equilibrium can be obtained by minimization of the potential energy (\ref{foo40}) under imposed Lees-Edwards boundary conditions \cite{allen1989computer}, and the nonaffine velocities are given by
\begin{equation}
\ket{\dot{\xv}} = -{\cal M} ^{-1} \ket{\partialbar_\gamma\partial_\xv U}\,,
\end{equation}
Elastic moduli in the athermal limit are given by
\begin{equation}
E \equiv \frac{1}{V} \frac{d^2 U}{d \gamma^2} =  \frac{1}{V} \left [ \frac{\partial^2 U}{\partial \gamma^2} -  \bra{\partialbar_\gamma\partial_\xv U } {\cal M}^{-1} \ket{ \partial_\xv\partialbar_\gamma U }\right] \,,
\end{equation}
a relation established in earlier work \cite{lutsko}.

\section{Framework for variations with respect to deformation}
\label{derivatives}
In this work we adopt a Lagrangian formulation, and express all variations with respect to the imposed deformations in terms of the \emph{deformed coordinates}. Deformations are imposed to our system by applying an affine transformation ${\cal H}(\gamma)$ --- parameterized by a strain parameter $\gamma$, as given e.g.~by Eqs.~(\ref{foo43}) and (\ref{foo44}) --- to the coordinates $\xv$, i.e.~$\xv\!\to\!{\cal H}\cdot\xv$. The coordinates' variations are supplemented by additional \emph{nonaffine displacements}, that are determined self-consistently by the geometric constraints embodied in the elastically-embedded frame of rigid edges (as expressed by Eq.~(\ref{foo00})), and by the mechanical equilibrium constraints (as expressed by Eq.~(\ref{foo02})). The total variation of pairwise differences $\xv_{ij}\!\equiv\!\xv_j\!-\!\xv_i$ follows
\begin{equation}\label{foo05}
\frac{d\xv_{ij}}{d\gamma} = \xv_{ij}\cdot\frac{d{\cal H}^T}{d\gamma} + \dot{\xv}_{ij}\,,
\end{equation}
where $\dot{\xv}$ denotes the nonaffine displacements per unit strain, referred to throughout our work as the \emph{nonaffine velocities}. We deliberately spell out the total variation of the pairwise differences $\xv_{ij}$ since in our systems we employ periodic boundary conditions. For simple shear deformations, we employ Lees-Edwards periodic boundary conditions \cite{allen1989computer}. Eq.~(\ref{foo05}) is the key relation that leads to a general form for the total variation with respect to the imposed deformation of any explicit function ${\cal Z}$ of the set of pairwise differences $\xv_{ij}$; it reads
\begin{eqnarray}
\frac{d {\cal Z}}{d\gamma} &= &\sum_{i<j}\frac{\partial {\cal Z}}{\partial\xv_{ij}}\cdot\frac{d\xv_{ij}}{d\gamma} \nonumber\\
&=&\sum_{i<j} \frac{\partial {\cal Z}}{\partial\xv_{ij}}\cdot\left(\frac{d{\cal H}}{d\gamma}\cdot\xv_{ij} + \dot{\xv}_{ij}\right)\,,
\end{eqnarray}
For the sake of brevity we define the operator
\begin{equation}\label{foo18}
\frac{\partialbar}{\partialbar \gamma} \equiv \sum_{i<j} \frac{\partial }{\partial\xv_{ij}}\cdot\frac{d{\cal H}}{d\gamma}\cdot\xv_{ij}\,,
\end{equation}
then the total variations with respect to deformation can be written in a compact form as
\begin{equation}
\frac{d}{d\gamma} = \frac{\partialbar}{\partialbar \gamma} + \dot{\xv}_k\cdot\frac{\partial}{\partial\xv_k}\,,
\end{equation}
where we have used that for any vector $\vv$
\begin{equation}
\vv_k\cdot\frac{\partial}{\partial\xv_k} = \sum_{i<j} \vv_k\cdot \frac{\partial \xv_{ij}}{\partial\xv_k}\cdot\frac{\partial}{\partial\xv_{ij}} = \sum_{i<j}\vv_{ij}\cdot\frac{\partial}{\partial\xv_{ij}}\,.
\end{equation}
In our work we consider explicit functions of the set of pairwise differences $\xv_{ij}$ \emph{and} of the set of edge forces $\tau_{ij}$; in these cases, the total variation reads
\begin{equation}
\frac{d}{d\gamma} = \frac{\partialbar}{\partialbar \gamma}+ \dot{\xv}_k\cdot\frac{\partial}{\partial\xv_k} + \dot{\tau}_{ij}\frac{\partial}{\partial\tau_{ij}}\,,
\end{equation}
where $\dot{\tau}$ denotes the variation of the edge forces with deformation. 

It is important to appreciate that the operator $\partialbar/\partialbar \gamma$ as defined in Eq.~(\ref{foo18}) does not commute with the spatial partial derivative $\partial/\partial\xv_k$; to see this, consider the variation of a pairwise distance with the imposed deformation
\begin{equation}
\frac{\partialbar r_{ij}}{\partialbar \gamma} = \frac{\xv_{ij}\cdot\frac{d{\cal H}^T}{d\gamma}\cdot\xv_{ij}}{r_{ij}}\,,
\end{equation}
then the spatial variation of the above follows as
\begin{eqnarray}
\frac{\partial}{\partial\xv_k}\frac{\partialbar r_{ij}}{\partialbar \gamma} & = & \left(\frac{\xv_{ij}\cdot\frac{d{\cal H}^T}{d\gamma} + \xv_{ij}\cdot\frac{d{\cal H}}{d\gamma}}{r_{ij}} \right. \nonumber\\
& & \quad - \left. \frac{\big(\xv_{ij}\cdot\frac{d{\cal H}^T}{d\gamma}\cdot\xv_{ij}\big)\xv_{ij}}{r_{ij}^3} \right)\cdot\frac{\partial\xv_{ij}}{\partial\xv_k} \label{foo23}\,,
\end{eqnarray}
On the other hand, the spatial derivative of a pairwise distance reads
\begin{equation}\label{foo24}
\frac{\partial r_{ij}}{\partial\xv_k} = \frac{\xv_{ij}}{r_{ij}}\cdot\frac{\partial\xv_{ij}}{\partial\xv_k}\,,
\end{equation}
with a variation with the imposed strain that follows
\begin{eqnarray}
\frac{\partialbar}{\partialbar \gamma}\frac{\partial r_{ij}}{\partial\xv_k} & = & \left(\frac{\xv_{ij}\cdot\frac{d{\cal H}^T}{d\gamma}}{r_{ij}}- \frac{\big(\xv_{ij}\cdot\frac{d{\cal H}^T}{d\gamma}\cdot\xv_{ij}\big)\xv_{ij}}{r_{ij}^3} \right)\cdot\frac{\partial\xv_{ij}}{\partial\xv_k} \nonumber\\
& \ne &\ \  \frac{\partial}{\partial\xv_k}\frac{\partialbar r_{ij}}{\partialbar \gamma}\,.
\end{eqnarray}
This non-commuting property of mixed variations has been overlooked in previous work, since for generic athermal elastic solids in mechanical equilibrium
\begin{equation}
\frac{\partialbar}{\partialbar\gamma}\frac{\partial U}{\partial\xv} = \frac{\partial }{\partial\xv} \frac{\partialbar U}{\partialbar\gamma}\quad\mbox{if}\quad\  \frac{\partial U}{\partial\xv} = 0\,,
\end{equation}
as acknowledged for instance in \cite{sum_rules}. In our work the potential alone may not satisfy mechanical equilibrium (see e.g.~Eq.~(\ref{foo01}), therefore the order in which the spatial and deformation-induced variations are considered is important.

\section{The kernel of ${\cal S}$ and its associated projection operator}
\label{projection_operator_appendix}

To show that Eq.~(\ref{foo08}) is a unique solution to the mechanical equilibrium equation (\ref{foo04}) we insert Eq.~(\ref{foo08}) into Eq.~(\ref{foo04}) and rearrange, to find
\begin{equation}\label{foo07}
\big({\cal I} - {\cal S}^T({\cal S}{\cal S}^T)^{-1}{\cal S}\big)\ket{\partial_\xv U} = \zerovector\,,
\end{equation}
with ${\cal I}$ denoting the identity operator. Therefore, we need to prove that ${\cal I}\!-\!{\cal S}^T({\cal S}{\cal S}^T)^{-1}{\cal S}$ is a projection operator onto the kernel of ${\cal S}$ \cite{states_of_self_stress_epje}.

The symmetric, positive semi-definite operators ${\cal S}{\cal S}^T$ and ${\cal S}^T{\cal S}$ share the same non-zero eigenvalues. In addition, their respective eigenvectors are related by a simple relation \cite{asm_pnas}; if ${\cal S}{\cal S}^T\ket{\phi_\omega}\!=\! \omega^2\ket{\phi_\omega}$ and ${\cal S}^T{\cal S}\ket{\Psi_\omega}\! =\! \omega^2\ket{\Psi_\omega}$ with the same $\omega$, then ${\cal S}\ket{\Psi_\omega}\! =\! \omega\ket{\phi_\omega}$ and ${\cal S}^T\ket{\phi_\omega}\! =\! \omega\ket{\Psi_\omega}$. We decompose the operator ${\cal I}\!-\!{\cal S}^T\!\left({\cal S}{\cal S}^T\right)^{-1}\!{\cal S}$ on the eigenmodes of ${\cal S}{\cal S}^T$ and ${\cal S}^T{\cal S}$ to find
\begin{eqnarray}
{\cal I}\!-\!{\cal S}^T({\cal S}{\cal S}^T)^{-1}{\cal S}  & = &  \sum_\omega\ket{\Psi_\omega}\bra{\Psi_\omega} - \sum_{\omega>0}\frac{{\cal S}^T\ket{\phi_\omega}\bra{\phi_\omega}{\cal S}}{\omega^2} \nonumber \\
& = & \sum_\omega\ket{\Psi_\omega}\bra{\Psi_\omega} - \sum_{\omega>0}\ket{\Psi_\omega}\bra{\Psi_\omega} \nonumber \\ 
& = & \sum_{\omega=0}\ket{\Psi_\omega}\bra{\Psi_\omega} \nonumber  \,. 
\end{eqnarray}
The zero-frequency modes $\ket{\Psi_{\omega}}$ of ${\cal S}^T{\cal S}$ form an orthonormal basis of the kernel of ${\cal S}$. In particular, any floppy mode $\ket{\uv}$ (i.e.~that solves ${\cal S}\ket{\uv}\!=\!0$) belongs to the kernel vector space.

\section{Numerical protocols, models and methods}
\label{numerics_appendix}

In this Appendix we provide a detailed description of the protocols used to generate the embedded frames of rigid edges, we describe the embedding potential energies of the frames considered in our work, and descibe the numerical method that derives from the theoretical framework developed in Sect.~\ref{theoretical_framework}. 

\subsection{Networks}
Simulations are performed on two types of 2D-networks. The first type of networks were constructed by first generating disordered packings of compressed bi-disperse disks, following the methods described e.g.~in \cite{breakdown}. We then obtain highly coordinated ($z\sim5-6$) contact networks from these packing. We next dilute the network of contacts by removing its edges, while aiming to preserve the homogeneity of the local coordination number of nodes. This is achieved by prefering the removal of edges from the highly coordinated particles. Using this protocol we created networks with various coordinations in the range [3-3.99]. An example of a network obtained using this protocol is shown in  Fig.~\ref{isotropic_mechanics}a. 

The second type of networks are off-lattice honeycomb networks. A full regular honeycomb network is created with a local and global coordination of $3$. We then dilute the networks by randomly removing edges with a probability~$(1-P_\text{bond})$. This random dilution probability $P_{\text{bond}}$ is used to control the resulting coordination $z$. However, its value is bounded by a lower limit set by percolation probability of the specific lattice (For honeycomb-lattice $P_{\text{bond}}>0.6527$ \cite{1964JMP}). After dilution, any rattlers or dangling ends are removed. Spatial disorder is then introduced by displacing the nodes in a random direction with a random magnitude between $0$ and $d_\text{max}$.  An example of a network obtained from this protocol is shown in  Fig.~\ref{isotropic_mechanics}b. 

The packing based networks allow us to probe mechanics over a wide range of coordinations, and, in particular, study phenomena that emerge close to $z_c$. The honeycomb networks allow sampling a smaller window of coordinations, quite far from $z_c$, but appear to be more robust against plastic instabilities, described in Appendix~\ref{plastic_instabilities}.

\subsection{Embedding elastic energy}
Having explained how we generated floppy frames of rigid edges, we next introduce the elastic energy in which our frames are then embedded in. In order to establish the generality of our theoretical framework and results, we chose and employed two different forms of the potential energy function $U(\xv)$, that depends on the nodes' coordinates $\xv$. We indeed show in what follows that our results do not depend on the specific choice of the potential energy. 

The first potential energy function we employed is meant to model bending interactions between pairs of edges that share a common node, with no other edges in between them; it reads
\begin{equation}\label{foo30}
U = \frac{\kappa}{2} \sum_{\mbox{\tiny triples }i,j,k} (\theta_{ijk}-\theta_{ijk}^{(0)})^2 \,,
\end{equation}
where the sum is understood to run over the relevant triples. $\theta_{ijk}$ is the angle formed between two edges that share a common node, and $\theta_{ijk}^{(0)}$ is the `rest-angle' of the said interaction. Before any deformation is imposed, we assume that all angles reside precisely at their associated rest-angles. 

We have also employed a potential energy that consists of a simple network of Hookean springs. Given our frame of rigid edges, we place a Hookean spring between all nearby nodes that are \emph{not} already connected by a rigid edge. The potential then reads
\begin{equation}\label{foo31}
U = \frac{\kappa}{2} \sum_{\mbox{\tiny neighbors }i,j} (r_{ij} - r_{ij}^{(0)})^2  \,,
\end{equation}
where $r_{ij}$ is the distance between the $i\th$ and $j\th$ nodes, and $r_{ij}^{(0)}$ is the restlength of the said interaction. Before any deformation is imposed, we assume that all pairs connected by a Hookean spring reside precisely at their associated rest-length. 

In both potential energies given by Eqs.~(\ref{foo30}) and (\ref{foo31}) there appears an stiffness scale $\kappa$; since it is the only energy scale in the system, it forms our microscopic units of energy, together with the characteristic length of an edge.

\subsection{Quasistatic deformation simulations}
\label{aqs}
We impose quasistatic deformation of our complex solids of elastically-embedded rigid-edge frames as follows; at each step we solve Eq.~(\ref{foo03}) iteratively using a conjugate gradient method to obtain the nonaffine velocities $\ket{\dot{\xv}}$ and the edge-force variations $\ket{\dot{\tau}}$. We then impose a small shear or dilatant strain increment $\Delta\gamma$ (as described in Appendix \ref{derivatives}), evolve the coordinates according to the linear approximation $\ket{\xv}\!\rightarrow\!{\cal H}(\Delta\gamma)\ket{\xv}\!+\!\Delta \gamma\ket{\dot{\xv}}$, and the edge forces according to $\ket{\tau}\!\rightarrow\!\ket{\tau}\!+\!\Delta\gamma\ket{\dot{\tau}}$. These steps are repeated while adjusting the strain increment such that $|\dot{\xv}|\Delta \gamma$ remains constant, until the strain stiffening transition is reached; we typically end our deformation when the strain to the stiffening transition is of order $10^{-5}$.  
 
The evolution of the network configuration by finite integrations steps will inevitably lead to a violation of the incompressibility/inextensibility of the edges, and mechanical equilibrium constraints on the frame's nodes. We however are able to bound the accumulated error by systematically performing correction steps in which an adjustment of the nodes' positions and of the edge forces restore the satisfaction of the said constraints. The formulation of the correction step is described next.

We first show that there exist a displacement of the nodes $\delta \xv$ and a correction of the edge forces $\delta \tau$ such that, when applied to a configuration, force balance is restored. This means that
\begin{equation}\label{afoo02}
({\cal S}^T + \delta{\cal S}^T)\ket{\tau + \delta \tau} - \ket{\sFrac{\partial U}{\partial \xv}\big|_{\xv + \delta \xv} } = 0\,.
\end{equation}
If the displacements $\delta \xv$ are small, we can approximate the forces at the new positions as
\begin{equation}\label{afoo03}
\ket{\sFrac{\partial U}{\partial \xv}\big|_{\xv + \delta \xv }} \simeq  \ket{\sFrac{\partial U}{\partial \xv}\big|_{\xv}} + {\cal M}\ket{\delta \xv}\,.
\end{equation}
The change in ${\cal S}^T$ due to the convection by $\delta\xv$ is expressed as
\begin{equation}\label{afoo04}
\delta{\cal S}^T\ket{\alpha} = \frac{\partial^2r_\alpha}{\partial \xv \partial \xv}\cdot \delta \xv\,.
\end{equation}
Using (\ref{afoo03}) and (\ref{afoo04}) in (\ref{afoo02}), we obtain
\begin{equation}
{\cal S}^T\ket{\tau} + {\cal S}^T\ket{\delta \tau} + \delta{\cal S}\ket{\tau} -\ket{\sFrac{\partial U}{\partial \xv}} - {\cal M}\ket{\delta \xv} =0\,,
\end{equation}
where we omitted terms of order $\delta \xv \delta \tau$.

In addition to bringing the system back to mechanical equilibrium, the displacement of the nodes $\delta\xv$ should also cancel the errors accumulated in the actual bar lengths, which means
\begin{equation}
({\cal S} + \delta{\cal S})\ket{\xv + \delta \xv} = \ket{\ell}\,,
\end{equation}
where we denoted the true lengths of the rods by $\ell_\alpha$. Using again the variation of ${\cal S}$, to first order in the displacement $\delta \xv$ the above relation becomes
\begin{equation}
{\cal S}\ket{\xv} + {\cal S}\ket{\delta \xv} + \delta{\cal S}\ket{\xv} = \ket{\ell}\,.
\end{equation}
Notice that ${\cal S}\ket{\xv} = \ket{r}$, and that
\begin{equation}
\bra{\alpha}\delta{\cal S}\ket{\xv} = \delta\xv\cdot\frac{\partial^2 r_\alpha}{\partial\xv\partial\xv}\cdot\xv = 0\,,
\end{equation}
and therefore Eq.~(\ref{foo13}) becomes
\begin{equation}
-{\cal S}\ket{\delta \xv} = \ket{r-\ell}\,.
\end{equation}

The correction step is therefore done by displacing the nodes according to $\ket{\xv}\!\rightarrow\!\ket{\xv}\!+\!\ket{\delta \xv}$, and varying the edge forces according to $\ket{\tau}\!\rightarrow\!\ket{\tau}\!+\!\ket{\delta \tau}$, where $\ket{\delta \xv}$ and $\ket{\delta \tau}$ are solutions to the equations
 \begin{equation}
\left( \begin{array}{cc}{\cal A}(\tau)&-{\cal S}^T\\-{\cal S}&0\end{array}\right)
\left( \begin{array}{c} \ket{\delta \xv} \\ \ket{\delta \tau} \end{array}\right) = 
\left( \begin{array}{c} \ket{\fv} \\\ket{r-\ell}\end{array}\right)\,,
\end{equation}
where $\ket{\fv}$ are the unbalanced net forces, and $\ket{r-\ell}$ are the differences between the current edge lengths and what their true lengths should be, both stemming from the accumulation of integration errors. The correction step described here can be repeated until the violation of the constraints becomes smaller than the desired precision. In our simulations we have chosen $10^{-8}$ as the bound on the accumulated relative error.

\begin{figure}[b]
\centering
\includegraphics[width = 0.5\textwidth]{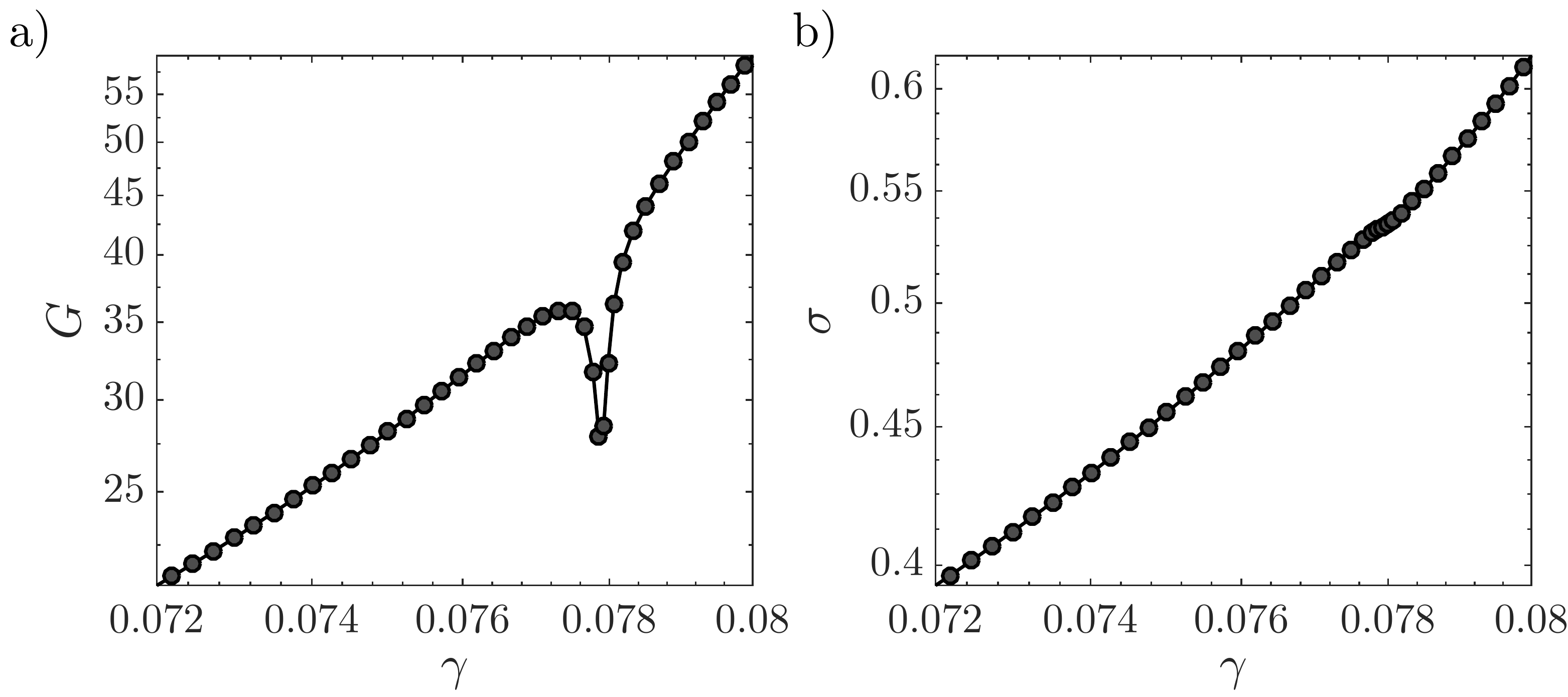}
\caption{\footnotesize Example of a plastic instability in a packing-derived network ($N=1600$, $z=3.8$). (a) The shear modulus shows a characteristic dip where the network softens over a small strain interval. (b) The shear stress shows a sign of the instability as well.}
\label{instability}
\end{figure}

\section{Plastic instabilities}
\label{plastic_instabilities}
Upon the quasistatic deformation of the elastically-embedded frames it is observed that transitions can occur between metastable states by local buckling events. When instabilities are visible, they manifest as soft-spot in the modulus and stress, as demonstrated in Fig.~\ref{instability} that shows the signature of a plastic instability in the shear modulus. The prevalence of these plastic events appears to depend on the coordination and system size, as well as on the details of the elastic interactions in which the frames are embedded. We leave the systematic investigation of these instabilities for future research. 

\vspace{0.1cm}
\section{Dominant term of $\ket{\dot{\tau}}$ close to $\gamma_c$}
\label{remainbound}

Using the approximation Eq.~(\ref{foo33}) in Eq.~(\ref{foo15}) for the edge force variations, one finds
\begin{equation}\label{foo42}
\ket{\dot{\tau}} \simeq \tau\frac{\braket{\phi_0}{\partialbar_\gamma r}}{\omega_0^2}\ket{\phi_0} + \tau\frac{\bra{\phi_0}{\cal S}{\cal A}^{-1}\ket{{\partialbar_\gamma \fv}}}{\omega_0^2}\ket{\phi_0}\,.
\end{equation}
Let us focus on the second term on the RHS of the above relation, and in particular on the contraction
\begin{equation}
\bra{\phi_0}{\cal S}{\cal A}^{-1}\ket{{\partialbar_\gamma \fv}} = \omega_0 \bra{\Psi_0}{\cal A}^{-1}\ket{\partialbar_\gamma\fv}\,.
\end{equation}
Since $\ket{\partialbar_\gamma\fv}$ and the matrix elements of $\cal A$ depend \emph{linearly} on the edge forces $\tau$, one could expect that $\bra{\partialbar_\gamma\fv}{\cal A}^{-2}\ket{\partialbar_\gamma\fv}$ remains finite as $\gamma\!\to\!\gamma_c$. This, in turn, implies that close to $\gamma_c$ we can neglect the second term on the RHS of Eq.~(\ref{foo42}), then 
\begin{equation}
\ket{\dot{\tau}} \simeq \tau\frac{\braket{\phi_0}{\partialbar_\gamma r}}{\omega_0^2}\ket{\phi_0}\,,
\end{equation}
as seen in Eq.~(\ref{foo41}). 

\section{The contraction $\bra{\tau}\partial_\xv\partialbar_\gamma r\ket{\dot{\xv}}$}
\label{robbiez_contraction}
In this Appendix we show that although both the edge forces $\tau$ and the nonaffine velocities $\dot{x}$ diverge as $\delta\gamma^{-1/2}$ upon approaching the strain stiffening transition, the contraction $\bra{\tau}\partial_\xv\partialbar_\gamma r\ket{\dot{\xv}}\!\sim\!\delta\gamma^{-1/2}$ and not $\sim\delta\gamma^{-1}$ as one might naively expect.

We start by using Eq.~(\ref{foo24}) in Eq.~(\ref{foo23}), to obtain
\begin{widetext}
\begin{equation}
\frac{\partial}{\partial\xv_k}\frac{\partialbar r_{ij}}{\partialbar \gamma}= \left(\frac{\xv_{ij}\cdot\frac{d{\cal H}^T}{d\gamma} + \xv_{ij}\cdot\frac{d{\cal H}}{d\gamma}}{r_{ij}} -\frac{\big(\xv_{ij}\cdot\frac{d{\cal H}^T}{d\gamma}\cdot\xv_{ij}\big)\xv_{ij}}{r_{ij}^3} \right)\cdot\frac{\partial\xv_{ij}}{\partial\xv_k} = \bigg(\frac{d {\cal H}}{d\gamma} + \frac{d{\cal H}^T}{d\gamma} + \frac{\xv_{ij}\cdot\frac{d{\cal H}^T}{d\gamma}\cdot\xv_{ij}}{r_{ij}^2}\bigg)\cdot\frac{\partial r_{ij}}{\partial\xv_k}
\end{equation}
The contraction of interest takes the form
\begin{eqnarray}
\sum_{\mbox{\tiny edges }i,j}\tau_{ij}\frac{\partial }{\partial\xv_k}\frac{\partialbar r_{ij}}{\partialbar\gamma}\cdot\dot{\xv}_k & = & \sum_{\mbox{\tiny edges }i,j}\tau_{ij}\bigg(\frac{d {\cal H}}{d\gamma} + \frac{d{\cal H}^T}{d\gamma} + \frac{\xv_{ij}\cdot\frac{d{\cal H}^T}{d\gamma}\cdot\xv_{ij}}{r_{ij}^2}\bigg)\cdot\frac{\partial r_{ij}}{\partial\xv_k}\cdot\dot{\xv}_k \nonumber\\
& = & \sum_{\mbox{\tiny edges }i,j}\tau_{ij}\bigg(\frac{d {\cal H}}{d\gamma} + \frac{d{\cal H}^T}{d\gamma} + \frac{\xv_{ij}\cdot\frac{d{\cal H}^T}{d\gamma}\cdot\xv_{ij}}{r_{ij}^2}\bigg)\cdot\frac{\partialbar r_{ij}}{\partialbar\gamma} \sim \tau  \sim \delta\gamma^{-1/2} \,.
\end{eqnarray}
\end{widetext}
where we have used that $\partialbar r_{ij}/\partialbar\gamma$ is regular, that
\begin{equation}
\frac{\partial r_{ij}}{\partial\xv_k}\cdot\dot{\xv}_k = \frac{\partialbar r_{ij}}{\partialbar\gamma}
\end{equation}
following Eq.~(\ref{foo00}), and recall that repeated coordinate indices are understood to be summed over.

%

\end{document}